\documentclass[12pt]{article}
\usepackage{newtxtext,newtxmath}
\usepackage{graphicx} 
\usepackage[letterpaper,margin=1in]{geometry} 
\renewenvironment{abstract}
	{\quotation}
	{\endquotation}
\date{}

\makeatletter
\renewcommand{\fnum@figure}{\textbf{Figure \thefigure}}
\renewcommand{\fnum@table}{\textbf{Table \thetable}}
\makeatother
\usepackage{scicite}
\usepackage{url}
\usepackage[utf8]{inputenc}
\usepackage[version=3]{mhchem}
\usepackage{bbding,enumitem}
\usepackage{pifont} 
\newcommand{\xmark}{\ding{55}}  

\usepackage{booktabs} 
\usepackage{float} 
\usepackage{gensymb} 
\def\scititle{New Crystal Structures Hide in Plain Sight:\\ A Stress Test for AI-Guided Materials Discovery}
\title{\bfseries \boldmath \scititle}
\author{
Xin Zhang,$^{1\dagger}$ Scott B. Lee,$^{1\dagger}$ Sudipta Chatterjee,$^{1\dagger}$ Hanqi Pi,$^{2}$ Yi Jiang,$^{2}$\and
Fatmag\"ul Katmer,$^{1}$ Emily G. Ward,$^{1}$ Daniel E. Widdowson,$^{3}$ Charles C. Tam,$^{4}$\and
Sarah Schwarz,$^{4}$ Connor J. Pollak,$^{1}$ Jaime M. Moya,$^{1}$ Grigorii Skorupskii,$^{1}$\and
Vitaliy A. Kurlin,$^{3}$ Stephen D. Wilson,$^{4}$ \and 
B. Andrei Bernevig,$^{5,2,6\ast}$, Leslie M. Schoop$^{1\ast}$\and
\small{$^{1}$Department of Chemistry, Princeton University, Princeton, New Jersey 08544, USA}\\
\small{$^{2}$Donostia International Physics Center (DIPC), 20018 San Sebasti\'an, Spain}\\
\small{$^{3}$Materials Innovation Factory, University of Liverpool, Liverpool L7 3NY, UK}\\
\small{$^{4}$Materials Department, University of California, Santa Barbara, California 93106, USA}\\
\small{$^{5}$Department of Physics, Princeton University, Princeton, New Jersey 08544, USA}\\
\small{$^{6}$IKERBASQUE, Basque Foundation for Science, Bilbao, Spain}\\
	\small$^\ast$Corresponding author. Email: lschoop@princeton.edu, bernevig@princeton.edu\and
	\small$^\dagger$These authors contributed equally to this work.
}
\begin{document} 
\maketitle
\newpage
\begin{abstract} \bfseries \boldmath
New types of crystal structures are discovered only rarely, and the artificial intelligence (AI) models now reshaping materials discovery have so far produced new chemical compositions within known structural families rather than genuinely new structures. We report \ce{GdNiSn4} and \ce{LuNiSn4}, intermetallics that adopt a previously unreported structure type, found not by computation but by exploratory synthesis. Single-crystal diffraction shows that the structure is an intergrowth of two known structural units. We then use this system as a benchmark for two leading generative models, \textit{MatterGen} and \textit{DiffCSP++}. For \textit{DiffCSP++}, the benchmark is performed in its crystallographically constrained setting, using the required space-group and Wyckoff-position inputs. Under our sampling budget, neither model recovers the experimentally reported monoclinic structure within the structural-matching tolerance. The generated structures are evaluated without further structural relaxation using the nonmagnetic analog \ce{LuNiSn4}, where we rule out $4f$ magnetism as the cause. Because the new structure is built from familiar building blocks, it should be derivable. We argue that encoding chemical reasoning, such as the stacking of known motifs, is a concrete path toward AI that can discover structurally novel materials.
\end{abstract}

\newpage
\section{Introduction}
The crystal structure of a material sets its properties, yet the number of distinct structure types is surprisingly small. Most were established decades ago, and genuinely new ones are rarely found. Over the past decade, large materials databases and machine-learning (ML) models have transformed how we search for new compounds, and artificial intelligence (AI) is widely expected to accelerate materials discovery~\cite{Saal2013MaterialsOQMD, Jain2013, Horton2025AcceleratedProject, Bradlyn2017TopologicalChemistry, Vergniory2019AMaterials, Draxl2019TheIntelligence, Haastrup2018TheCrystals, Zhou20192DMatPediaApproaches, Choudhary2020TheDesign, Zeni2025ADesign, Merchant2023, Szymanski2023}. These methods have been better at proposing new chemical compositions inside known structural families than at finding new structure types. The reason is structural. Most generative models are trained on existing databases, such as the Materials Project (MP), the Inorganic Crystal Structure Database (ICSD), and Alexandria-derived sources~\cite{Allmann2007TheICSD, cavignac2025ai}, which are dominated by common structures with small unit cells. A model that learns this distribution naturally reproduces it, returning substitutional variants of what it has already seen, in which one element replaces another on a known lattice~\cite{PRXEnergy.3.011002, Cheetham2024ArtificialDiscovery}.

This limitation matters when the goal is a new structure type rather than a new composition. A compound is compositionally new if it has a previously unreported formula, even in a known structure type; it is structurally novel if it has no adequate precedent among known structure types. Establishing this distinction is not reducible to space group and Wyckoff sequence alone. In the ICSD framework, structure-type assignment begins with isopointal descriptors such as space group and Wyckoff sequence, but additional crystallographic descriptors are needed to subdivide isopointal structures into isoconfigurational structure types~\cite{Allmann2007TheICSD}. We use the Pointwise Deviation from Asymptotic (PDA) distance~\cite{widdowson2022resolving, widdowson2025geographic, widdowson2026pointwise} as a screen for structural redundancy. In this work, PDA distance is used to make the structural novelty workflow faster by eliminating near-duplicates before crystallographic comparison. This matters in practice, because nominally new candidates from high-throughput generative models have been shown to be near-duplicates of structures already in the ICSD and the MP~\cite{widdowson2025geographic, widdowson2026pointwise}.

Predicting whether a proposed structure can be synthesized is equally hard. Stability is generally judged from static, 0~K density functional theory (DFT) metrics such as the energy above the convex hull, yet these metrics are often insufficient for assessing experimental synthesizability~\cite{Bartel2022ReviewSolids}. Rare-earth intermetallics are a particularly demanding case, where their localized $4f$ electrons and strong spin-orbit coupling produce rich magnetism and correlated physics~\cite{Stewart1984Heavy-fermionSystems, Fisk1988Heavy-ElectronMatter, Yu2010Real-spaceCrystal, Skorupskii2024DesigningSemimetals}, but also make the underlying electronic-structure calculations expensive and less reliable \cite{Sholl2009DensityIntroduction, Singh2022Machine-learningCompounds}. The same ingredients that make these materials interesting also make them hard to predict. 

This difficulty does not mean structure prediction is intrinsically ineffective. Rather, it highlights the importance of incorporating chemically meaningful constraints. Existing successes in chemically constrained structure-prediction models clarify which priors enable reliable structure generation. Structure prediction has worked well in the realm of metal-organic frameworks and hybrid organic-inorganic perovskites, where known building blocks and connectivity rules constrain the structure-search space to be more tractable~\cite{Xu2023ExperimentallyMaterials, Karimitari2024AccuratePerovskites}. These successes point to a key ingredient: crystallochemical priors. These priors are derived from the intuition that solid-state chemists use to anticipate which structures can form, including electron-counting rules, preferred coordination environments, atomic packing constraints, and recurring structural motifs. This intuition can easily be encoded in the form of descriptors or constraints. Incorporating such priors into material prediction workflows would allow structure generation to move beyond reproducing patterns in existing databases and toward the prediction of structurally novel materials. 

Here, we report \ce{GdNiSn4} and \ce{LuNiSn4}, two rare-earth intermetallics that adopt a previously unreported structure type, which we identified through exploratory synthesis rather than computational prediction. Single-crystal X-ray diffraction (SCXRD) shows that both crystallize in the same monoclinic structure, which can be described as an intergrowth of two known structural motifs: a \ce{ZrGa2}-type \ce{RSn2} (R = Gd, Lu) slab and a \ce{NiSn2} slab attributed to the polytypic \ce{PdSn2}/\ce{CoGe2} structure types. This reassigns the previously reported orthorhombic \ce{LuNiSn4} model~\cite{Skolozdra2000NewProperties} to a monoclinic structure type, a correction that has already propagated into materials prediction workflows, where several \ce{RNiSn4} phases appear in the orthorhombic structure while \ce{GdNiSn4} was not predicted at all in the Materials Project database~\cite{Jain2013}. We explain the monoclinic preference using density functional theory (DFT) calculations and chemical-pressure arguments. We also use the nonmagnetic analog \ce{LuNiSn4} to show that $4f$ magnetism is not responsible for this structural preference. We then use this system as a benchmark of two leading generative models, \textit{MatterGen}~\cite{Zeni2025ADesign} and \textit{DiffCSP++}~\cite{jiao2024space}, neither of which recovers the monoclinic structure without strong crystallographic constraints. Finally, magnetic, electrical transport, and heat-capacity measurements reveal complex magnetism in \ce{GdNiSn4}, whose ground state we have recently resolved as a collinear spin density wave~\cite{GdNiSn4_SDW}. Because the new structure is assembled from familiar building blocks, it should in principle be derivable, and we argue that teaching models to reason about the stacking of known motifs is a practical path toward AI that can discover new structure types.

\section*{Results}
\subsection*{Identification of an unreported structure type}
Millimeter-scale single crystals of \ce{GdNiSn4} were grown by the self-flux~\cite{Canfield01061992} method~(Fig.~\ref{GdniSn4_Crystals}) and characterized by SCXRD and SEM-EDS~(Fig.~\ref{GdniSn4_SEM}). \ce{GdNiSn4} crystallizes in the centrosymmetric monoclinic space group $C2/m$~(No.~12)~(Tables~\ref{R_114}--\ref{U_APP}). Gd occupies two crystallographically distinct $4i$ sites, and Ni occupies one $8j$ site. Sn is distributed in two $8j$ sites, three $4i$ sites, and one $4h$ site. The resulting Wyckoff sequence is $j3i5h$. Because most atoms occupy non-maximal Wyckoff positions with free internal coordinates, the structure is sensitive to positional parameters, so a shared space group and Wyckoff sequence by itself does not imply the same arrangement of atoms. 

\begin{figure}[H] 
\centering
\includegraphics[width=1.0\textwidth]{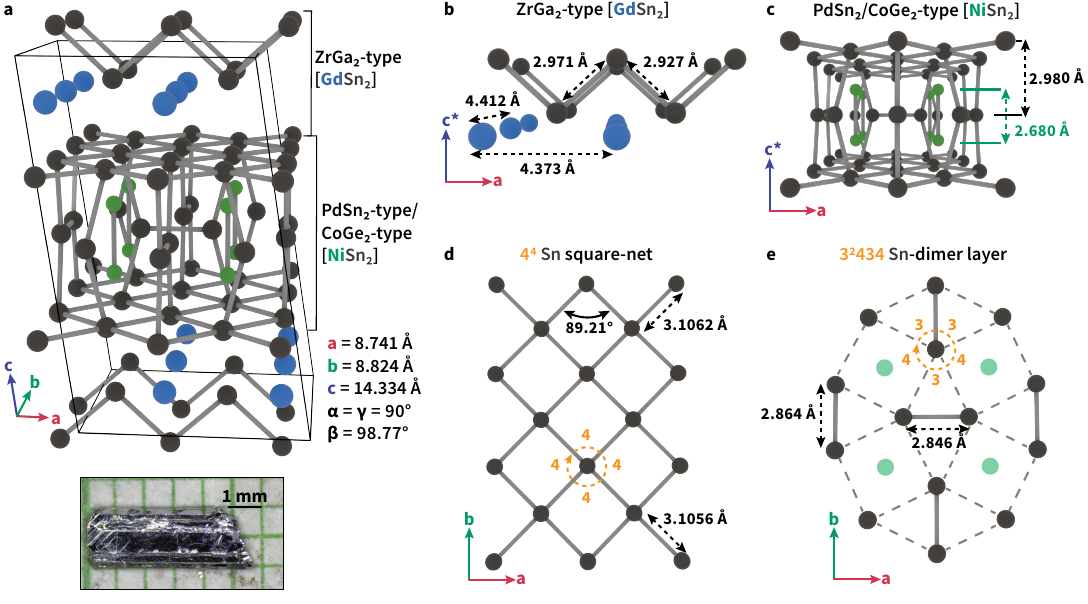}
\caption{\textbf{Monoclinic \ce{GdNiSn4} crystal structure.} \textbf{(a)} The overall \ce{GdNiSn4} structure ($C2/m$, No.~12)  composed of alternating \ce{GdSn2} and \ce{NiSn2} units. The inset shows a single crystal of \ce{GdNiSn4}. \textbf{(b)} The \ce{ZrGa2}-type \ce{GdSn2} unit consists of uncoordinated Gd atoms occupying sites adjacent to the Sn zigzag chains that run along~$a$, forming a corrugated \ce{GdSn2} sheet. \textbf{(c)} In-plane projection down $b$ of the \ce{PdSn2}-type/\ce{CoGe2}-type \ce{NiSn2} unit, highlighting the connectivity between the $4^{4}$ Sn square-net and $3^{2}434$ Sn-dimer sublayers. The Sn-dimer sublayer links the square-net sublayers above and below and is interleaved with an array of Ni dimers. \textbf{(d)} Projection down $c^*$ (normal to $ab$) showing the $4^{4}$ Sn square-net sublayer. \textbf{(e)} Projection down $c^*$ showing the $3^{2}434$ Sn-dimer sublayer. Dashed lines help visualize the $3^{2}434$ vertex configuration. Bonds are drawn for all contacts shorter than 3.15$\,$\AA. In panels where axes are labeled $a$, $b$, and $c^*$, $c^*$ denotes the reciprocal-lattice vector normal to the $ab$-plane.}
\label{Gd114_S}
\end{figure}

The structure of \ce{GdNiSn4}~(Fig.~\ref{Gd114_S}a) is built from alternating \ce{GdSn2} and \ce{NiSn2} units. The \ce{GdSn2} unit, of the \ce{ZrGa2} structure type, consists of uncoordinated Gd sites adjacent to Sn zigzag chains running along $a$~(Fig.~\ref{Gd114_S}b) that form a corrugated \ce{GdSn2} sheet. The \ce{GdSn2} unit alternates with a \ce{NiSn2} unit related to both the polytypic \ce{PdSn2} and \ce{CoGe2} structure types, which differ only in stacking sequence~\cite{Leineweber2021Preparation10GPa}. The \ce{NiSn2} unit consists of two $4^{4}$ Sn square-net sublayers sandwiching a $3^{2}434$ Sn-dimer sublayer, with Ni dimers connecting above and below the four-membered rings of the $3^{2}434$ net~(Fig.~\ref{Gd114_S}c--e). We describe individual atomic layers using the vertex configuration notation, where the integers indicate the sizes of the polygonal rings meeting at each vertex (atom), and the superscripts denote the multiplicity of each ring size. For example, $4^{4}$ denotes four squares meeting at each vertex, whereas $3^{2}434$ corresponds to the sequence 3--3--4--3--4 around a vertex (three triangles and two squares). This vertex configuration matches the Shastry-Sutherland lattice model\cite{SriramShastry1981ExactAntiferromagnet}. The short spacing between the Sn square-net and Sn-dimer sublayers indicates that these motifs form a single connected Sn framework rather than structurally isolated layers, consistent with what is seen in the \ce{PdSn2} and \ce{CoGe2} structure types. 

To test whether this arrangement is structurally novel, we did not rely on the space group and Wyckoff sequence alone. The ICSD lists two other $C2/m$ entries with the same Wyckoff sequence, \ce{CdVO(SeO3)2}~(\#209150) and \ce{K2In(PS4)(P2S6)_{0.5}}~(\#248034), so a match at that level would be misleading. We therefore searched the ICSD and the Materials Project with the PDA distance, which compares the actual periodic geometries~\cite{widdowson2022resolving, widdowson2026pointwise}. The ICSD/MP PDA approach used here~\cite{widdowson2025geographic} was run on a standard desktop computer, so this type of structural audit is not a computational bottleneck and can easily be utilized as a post-generation filter before AI-generated candidates are claimed to be structurally novel. Table~\ref{PDA} lists the closest entries. The nearest neighbors are all the reported \ce{RNiSn4} phases, despite being indexed in a different (orthorhombic) space group, whereas the two isopointal $C2/m$ entries lie far away by PDA distance. The structures that share the space group and Wyckoff sequence of \ce{GdNiSn4} are not its closest relatives, and \ce{GdNiSn4} has no isoconfigurational analog in either database. The comparisons support assigning \ce{GdNiSn4} to a previously unreported structure type. 

\begin{table}[H]
\centering
\caption{Eleven structures from the ICSD and Materials Project (MP) with the smallest PDA distance to \ce{GdNiSn4}\cite{widdowson2025geographic}. The closest entries differ by rare-earth element substitution in the \ce{RNiSn4} family, with \ce{Dy2NiSn6} also differing in stoichiometry. The two entries that share the same Wyckoff sequence and space group have the largest PDA distance to \ce{GdNiSn4}, indicating that they are not close isoconfigurational analogues.}
\label{PDA}
\vspace{6pt}
\begin{tabular}{lccc}
\hline
Composition                &Space Group         &Entry ID         &PDA distance (\AA) \\
\hline
\ce{LuNiSn4}               & $Cmmm$ (No.~65)    & MP-31433        & 0.0359            \\
\ce{TbNiSn4}               & $Cmmm$ (No.~65)    & MP-1208362      & 0.0644            \\
\ce{ErNiSn4}               & $Cmmm$ (No.~65)    & MP-1212981      & 0.0681            \\
\ce{HoNiSn4}               & $Cmmm$ (No.~65)    & MP-1212090      & 0.0831            \\
\ce{DyNiSn4}               & $Cmmm$ (No.~65)    & MP-1212896      & 0.0964            \\
\ce{TmNiSn4}               & $Cmmm$ (No.~65)    & MP-1207718      & 0.0976            \\
\ce{LuNiSn4}               & $Cmmm$ (No.~65)    & ICSD-106923     & 0.1363            \\
\ce{Dy2NiSn6}              & $Cmmm$ (No.~65)    & MP-1079605      & 0.4975            \\
\ce{Dy2NiSn6}              & $Cmmm$ (No.~65)    & ICSD-165783     & 0.4996            \\
\ce{K2In(PS4)(P2S6)_{0.5}} & $C2/m$ (No.~12)    & ICSD-248034     & 0.7480            \\
\ce{CdVO(SeO3)2}           & $C2/m$ (No.~12)    & ICSD-209150     & 1.0155            \\
\hline
\end{tabular}
\end{table}

\subsection*{Correction of the reported \ce{LuNiSn4} structure}
\begin{figure}[H]
\centering
\includegraphics[width=1.00\textwidth]{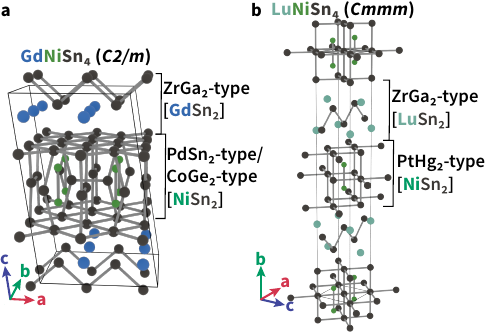}
\caption{\textbf{\ce{GdNiSn4} and \ce{LuNiSn4} crystal structures.} \textbf{(a)} \ce{GdNiSn4} crystallizes in the monoclinic space group $C2/m$~(No.~12) and is described as an alternation of \ce{ZrGa2}-type \ce{GdSn2} units and \ce{PdSn2}/\ce{CoGe2}-type \ce{NiSn2} units. \textbf{(b)} The previously reported\cite{Skolozdra2000NewProperties} \ce{LuNiSn4} model crystallizes in the orthorhombic space group $Cmmm$ (No.~65) and is described as an alternation of \ce{ZrGa2}-type \ce{LuSn2} units and \ce{PtHg2}-type \ce{NiSn2} units. Our structure refinement shows that \ce{LuNiSn4} instead adopts the same structure type as \ce{GdNiSn4}. The orthorhombic structure is therefore shown here as a higher symmetry reference.  Bonds are drawn for all contacts shorter than 3.15$\,$\AA.}
\label{Gd_Lu_Comp}
\end{figure}

The closest neighbors in Table~\ref{PDA} are the reported \ce{RNiSn4} phases, which were assigned to an orthorhombic \ce{LuNiSn4} structure type~($Cmmm$, No.~65) \cite{Skolozdra2000NewProperties}. In this model, the \ce{NiSn2} unit is built from three $4^{4}$ Sn square-net layers that form a \ce{PtHg2}-type unit~\cite{Bauer1953RontgenographischePlatin-Quecksilber}, rather than the two square nets plus a $3^{2}434$ Sn-dimer sublayer that we find. The original report itself noted warning signs: an unusually long orthorhombic axis ($b = 55.5$~\AA), supercell indexing, partial Sn occupancies near 50\%, and nonphysical Sn-Sn distances of 1.0 to 1.5~\AA. All this is explained if the middle square net is a Sn-dimer sublayer, because modeling dimerized Sn as an ideal square net places spurious electron density about 1 to 1.5~\AA\ from the true atomic sites. To resolve this ambiguity, we grew single crystals of \ce{LuNiSn4} and solved its structure~(Section~\ref{SCXRD}). \ce{LuNiSn4} adopts the same monoclinic structure type as \ce{GdNiSn4} (Tables~\ref{R_114_LuNiSn4}--\ref{U_APP_LuNiSn4}, Fig.~\ref{Gd_Lu_Comp}). The previously reported orthorhombic model is therefore an averaged, higher-symmetry description that does not capture the true monoclinic structure. Single-crystal indexing of \ce{RNiSn4} for R = Tb, Dy, Ho, Er, and Y reveals the same monoclinic cell, indicating that the monoclinic type is likely general to the family, although full refinements of additional members will be needed to establish this and will be reported separately.

This correction has practical consequences for materials prediction. The orthorhombic assignment has already propagated into the Materials Project, which lists several \ce{RNiSn4} phases in the orthorhombic type (Table~\ref{PDA}) while not predicting \ce{GdNiSn4} at all \cite{Jain2013}. Notably, the reported orthorhombic \ce{LuNiSn4} lies farther from the monoclinic \ce{GdNiSn4} structure by PDA than the database \ce{RNiSn4} entries do. Because Materials Project structures are DFT-relaxed within the symmetry of their starting model, this suggests that the atoms relax part-way toward the monoclinic structure but cannot reach it while constrained to the orthorhombic space group. An incorrect experimental assignment can therefore become entrenched in materials databases, propagating erroneous predictions across an entire structural family. 

\subsection*{Why the monoclinic structure forms}
Having established the structure, we asked why it is preferred, using both electronic and packing arguments. Across every comparison we tested, DFT finds the monoclinic structure lower in energy than the orthorhombic alternative, by 36 to 77 meV/atom for both \ce{GdNiSn4} and \ce{LuNiSn4}~(Table~\ref{ECOMP}). This small, but not negligible preference survives changes of functional (PBE and the meta-GGA RSCAN), $\mathbf{k}$-mesh, and spin-orbit coupling, and it persists in spin-polarized ferromagnetic and antiferromagnetic calculations and across a $U' = 0$ to $8$~eV range for the Gd $4f$ states~(Section~\ref{DFTUSI}). Magnetic order changes the size of the energy difference but not its sign. Crucially, the nonmagnetic analog \ce{LuNiSn4} shows the same monoclinic preference, so the effect is structural and electronic, not magnetic. The same separation between structural stabilization and magnetic classification appears in convex hull calculations. They place monoclinic \ce{GdNiSn4} on the convex hull ($E_{\mathrm{hull}} = 0$~meV/atom) and monoclinic \ce{LuNiSn4} only 2.9~meV/atom above it. Thus, the monoclinic structure type is thermodynamically stable even without $4f$ magnetism. Magnetic order is needed to classify \ce{GdNiSn4} correctly on the hull, but it is not the general stabilizing mechanism for the structure type itself. 

\begin{table}[H]
\centering
\caption{\textbf{Energy comparison (eV/atom) between orthorhombic \textit{(Cmmm)} and monoclinic~\textit{(C2/m)} structures for \ce{GdNiSn4} and \ce{LuNiSn4}.} Bold energy values indicate the lower-energy structure for a given composition and calculation setting. All calculations were performed on fully relaxed structures, and used a 650~eV plane-wave cutoff. Each calculation type varied the exchange-correlation functional (XC), $\mathbf{k}$-point mesh size, spin-orbit coupling (SOC), whether $f$ electrons were placed in the valence ($f$-elec = \Checkmark) or core ($f$-elec = \xmark) shells, and spin-polarization (SP) with either a ferromagnetic (FM) or antiferromagnetic (AFM) collinear magnetic state. The final two rows indicate DFT+U calculation results, where U' = 6 eV in the Dudarev formalism in accordance with previous literature~\cite{dudarev_dftu_1998,petersen_gdu_2006}.}
\vspace{6pt}
\label{ECOMP}
\begin{tabular}{ccccccccc}
\toprule
&  &  &  & & \multicolumn{2}{c}{\ce{GdNiSn4} (eV/atom)} & \multicolumn{2}{c}{\ce{LuNiSn4} (eV/atom)} \\
\cmidrule(lr){6-7} \cmidrule(lr){8-9}
 XC &  $\mathbf{k}$-mesh &  SOC &  $f$-elec & SP & $Cmmm$ & $C2/m$ & $Cmmm$ & $C2/m$ \\
\midrule
RSCAN & $15^3$ & \xmark & \xmark & \xmark & -42.420 & \textbf{-42.486}&  -44.542 & \textbf{-44.599}\\
PBE & $15^3$ & \Checkmark& \xmark & \xmark & -4.655 & \textbf{-4.726} & -4.576 & \textbf{-4.651}\\
PBE & $9^3$ & \Checkmark & \xmark &\xmark & -4.654& \textbf{-4.726} & -4.575 & \textbf{-4.651}\\
PBE & $9^3$ & \xmark\ & \Checkmark\ &\xmark & -4.504 & \textbf{-4.563}  & -4.485 & \textbf{-4.562} \\
PBE+U & $9^3$ & \xmark\ & \Checkmark & \Checkmark -- FM & -5.645 & \textbf{-5.720} & N/A & N/A\\
PBE+U & $9^3$ & \xmark\ & \Checkmark & \Checkmark -- AFM & -5.683 & \textbf{-5.719} & N/A & N/A\\
\bottomrule
\end{tabular}
\end{table}

\begin{figure}[H]
\centering
\includegraphics[width=0.75\textwidth]{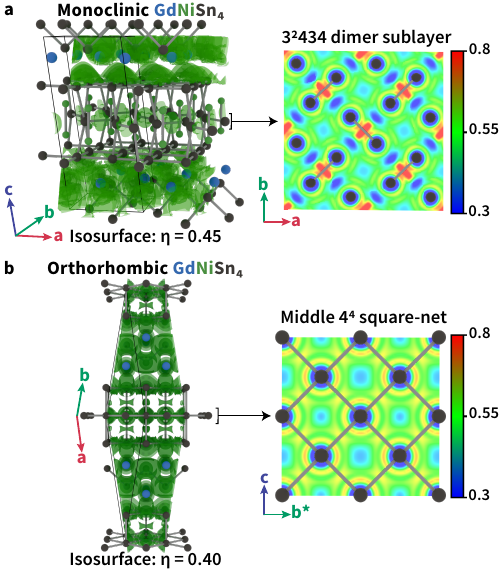}
\caption{\textbf{Electron localization function (ELF) isosurfaces and in-plane cutouts for monoclinic and orthorhombic \ce{GdNiSn4}.} \textbf{(a)} Monoclinic \ce{GdNiSn4} with an in-plane isosurface cut through the $3^{2}434$ Sn-dimer sublayer. The Sn zigzag chain is delocalized along $a$, and the connected $4^{4}$ Sn square-nets are delocalized along the $ab$ plane. Comparatively, the $3^{2}434$ Sn-dimer sublayer exhibits enhanced localization between the Sn--Sn dimers. \textbf{(b)} Orthorhombic \ce{GdNiSn4} with an in-plane isosurface cut through the middle $4^{4}$ square-net layer. The Sn zigzag chain along with the top/bottom $4^{4}$ square-nets, is delocalized in the same way as the monoclinic structure. The middle $4^{4}$ square-net exhibits enhanced localization between the Sn--Sn bonds in comparison to the top/bottom $4^{4}$ square-nets, but substantially weaker than in the monoclinic $3^{2}434$ Sn-dimer sublayer. All ELF results shown here were computed in the primitive cells of the monoclinic/orthorhombic structures.}
\label{ELF}
\end{figure}

What stabilizes the monoclinic form is the dimerization of Sn in the central $3^{2}434$ sublayer. An electron-localization function (ELF) analysis~(Fig.~\ref{ELF}) shows enhanced localization between the Sn atoms of the $3^{2}434$ sublayer, the signature of covalent Sn-Sn bonding, which is absent from the corresponding ideal square net of the orthorhombic model. Because the rendered isosurfaces can look different at different levels, we base this comparison on in-plane cross-sections plotted on the same ELF scale (An equivalent isosurface comparison is shown in in~Fig.~\ref{ELF_SI_fig}.) Outside this central layer the two structures are effectively identical, isolating Sn-Sn dimerization as the electronic driver.

A complementary steric argument explains why this particular intergrowth is favored. The \ce{NiSn2} unit is known as a binary that crystallizes in the \ce{CoGe2} structure type only after high-pressure treatment at 10~GPa~\cite{Leineweber2021Preparation10GPa}. In the related \ce{PdSn2} structure type, this motif is instead favored because alternating $4^{4}$ and $3^{2}434$ nets relieve the packing frustration that a structure of pure square nets or pure dimer nets would suffer from~\cite{Lim2024NavigatingClustering}. Inserting the \ce{ZrGa2}-type \ce{GdSn2}~slab supplies the same geometric relief that high pressure provides in the binary: the corrugated~\ce{GdSn2} interface expands the contacts that are over-compressed in binary~\ce{NiSn2}, so the high-pressure motif becomes tolerable at ambient pressure inside the ternary. The same packing logic fixes the arrangement between the slabs, requiring the \ce{GdSn2} unit above a \ce{NiSn2} unit to be shifted by $\tfrac{1}{4}a$ relative to the one below, which is what makes the cell monoclinic. The structure is thus not an arbitrary new arrangement but the predictable consequence of stacking two known units under a packing constraint.

\subsection*{Benchmark of AI structure prediction}
With the structure established, we used monoclinic \ce{GdNiSn4} and \ce{LuNiSn4} as a benchmark for two leading diffusion-based generative models, \textit{MatterGen}~\cite{Zeni2025ADesign} and \textit{DiffCSP++}~\cite{jiao2024space}. We used models trained on MPTS-52, since the 48-atom conventional cells of the experimental structures exceed the 20-atom limit of the common MP-20/Alex-MP-20 benchmarks~(Section~\ref{AI_SI_1}--\ref{AI_SI_2}). In a standard composition-conditioned task, the model is given the formula and generates the lattice and atomic positions. We compared each candidate to the experimental structure with the \textsc{pymatgen} \texttt{StructureMatcher} at three tolerance levels (Section~\ref{AI_SI_2}--\ref{AI_SI_3}). Two controls make the test interpretable. First, \ce{GdNiSn4} is absent from the MPTS-52 training set used here, so it is a genuinely held-out target, whereas \ce{LuNiSn4} appears once, but in the incorrect orthorhombic assignment. 

In the standard composition-conditioned \textit{MatterGen} task, no generated candidate matched either experimental target among 10,000 samples for each composition~(Section~\ref{AI_SI_4}). We then tested \textit{DiffCSP++} with substantially stronger crystallographic priors by fixing the composition, atom counts, space group, and experimentally realized Wyckoff template. Even with the experimental template supplied, \textit{DiffCSP++} produced no default-tolerance matches; only loose-tolerance near matches were obtained, with 6 loose and 10 very-loose matches for \ce{GdNiSn4}, and 13 loose and 18 very-loose matches for \ce{LuNiSn4} out of 10,000 samples each~(Section~\ref{AI_SI_5}). These results should not be interpreted as proof that the models can never generate the target structures or as a general failure of structure prediction. Rather, within the present sampling budget, the monoclinic \ce{RNiSn4}~(R = Gd, Lu) structures appear to occupy low-probability regions of the tested generative distributions, especially when only composition is provided.

The \textit{DiffCSP++} results show that near matches become accessible only after imposing strong crystallographic constraints. This highlights a practical limitation of current general-purpose AI generative models. Rare or sparsely represented structure types, particularly large-unit-cell intergrowths or superstructures, would require explicit structural-motif-level priors and more representative training data before they can be reliably rediscovered from composition or substitutional analogy alone. The similar outcome for magnetic \ce{GdNiSn4} and nonmagnetic \ce{LuNiSn4} argues that the difficulty is not an artifact of the Gd $4f$ electrons, but is associated with the large unit-cell, motif-stacked structure type itself. This problem is different from the substitutional material outcomes generative models have, and it is not reduced to the genetic algorithm searches already used in evolutionary CSP, which can recombine spatial fragments from parent structures~\cite{Glass2006USPEXEvolutionaryPrediction, Trimarchi2007GlobalConstraints}. Those methods can generate candidate intergrowths. However, they do not by themselves supply a packing-aware chemical prior that selects the viable stacking relation before candidate generation. In \ce{RNiSn4}, the missing information is not only which structural motifs are combined, but how those motifs are stacked and laterally offset so as to relieve local steric frustration while preserving the bonding motif that stabilizes the monoclinic structure. Reliable recovery of such structures will therefore require structural-motif-level priors, more representative large-cell training data, and post-generation physical validation, rather than composition-only generation or substitutional analogy alone. A fully characterized experimental counterexample makes that gap concrete and, as we discuss in more detail below, points to how it could be closed.

\subsection*{Complex magnetism in \ce{GdNiSn4}}
\begin{figure}[H]
\centering
\includegraphics[width=0.7\textwidth]{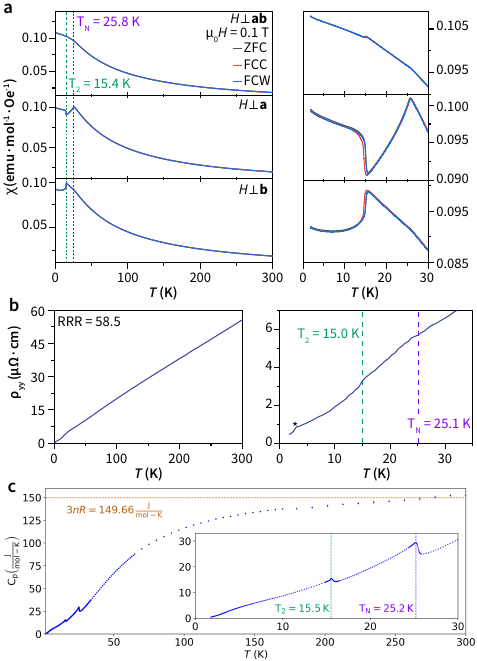}
\caption{\textbf{Temperature-dependent magnetic susceptibility, zero-field resistivity, and zero-field heat capacity of \ce{GdNiSn4}.} \textbf{(a)} Measurements were performed under zero-field cooled (ZFC), field-cooled cooling (FCC), and field-cooled warming (FCW) conditions for 3~orientations: $H\perp ab$ (top), $H\parallel a$ (middle), $H\parallel b$ (bottom). Two antiferromagnetic (AFM) transitions are observed at~$T_{\mathrm{N}}=25.8$~K and~$T_{\mathrm{2}}=15.4$~K with a FCC--FCW splitting of $\approx1$~K near $T_{\mathrm{2}}$ for all orientations. \textbf{(b)} Resistivity was performed with current applied along $b$~($\rho_{yy}$, $I \parallel y$), showing anomalies near $T_{\mathrm{N}}$ and $T_{\mathrm{2}}$. The residual resistivity ratio is calculated as~$\mathrm{RRR}=\rho(300\,\mathrm{K})/\rho(4\,\mathrm{K})$. The asterisk indicates a superconducting transition attributed to residual tin flux. \textbf{(c)} Heat capacity measurements exhibit two anomalies at T$_{N}$ = 25.2~K and T$_{2}$ = 15.5~K~(horizontal line indicates the Dulong-Petit limit).
}
\label{GdniSn4_MvT_RvT}
\end{figure}

A new structure type is most valuable if it also hosts interesting physics. Temperature-dependent susceptibility on an oriented \ce{GdNiSn4} single crystal reveals two antiferromagnetic-like transitions, at $T_{\mathrm{N}} = 25.8$~K and $T_2 = 15.4$~K, for all three field orientations, with the in-plane response ($H \parallel a$, $H \parallel b$) sharper than the out-of-plane response~(Fig.~\ref{GdniSn4_MvT_RvT}a). There is no observable thermal hysteresis at $T_{\mathrm{N}}$, whereas a small FCC--FCW splitting of $\approx1\,\mathrm{K}$ is observed around $T_{\mathrm{2}}$ for all orientations, consistent with a first-order phase transition. A Curie--Weiss fit~(Fig.\ref{GdniSn4_CW}) yields $\mu_{\mathrm{eff}}=7.93\,\mu_{\mathrm{B}}$, closely matching the $\mathrm{Gd}^{3+}$ free-ion value of $7.94\,\mu_{\mathrm{B}}$\cite{Mugiraneza2022Tutorial:Law}. This suggests that the magnetic response is dominated by localized Gd moments and is consistent with a magnetically inactive Ni site. This conclusion is consistent with neutron diffraction studies performed on related \ce{R-Ni-Sn}-based stannides~\cite{Romaka2002MagneticCompounds, Gil2003MagneticCompounds, Beirne2002MagneticNdNiSn, Yakinthos1995TheTbNiSn, Szytula1996MagneticTbRhSn, Yakinthos1994Amplitude-modulatedHoNiSn}. The fully occupied Ni $3d$ states in the \ce{GdNiSn4} projected density of states~(Fig.~\ref{fig:SIBandsnDos_EarlyVersion}) further supports this conclusion. Resistivity measurements show that \ce{GdNiSn4} is metallic with clear anomalies at both transitions~(Fig.~\ref{GdniSn4_MvT_RvT}b). Heat capacity measurements exhibit the same two corresponding anomalies, confirming that the transitions are bulk~(Fig.~\ref{GdniSn4_MvT_RvT}c).

\begin{figure}[H]
\centering
\includegraphics[width=1.0\textwidth]{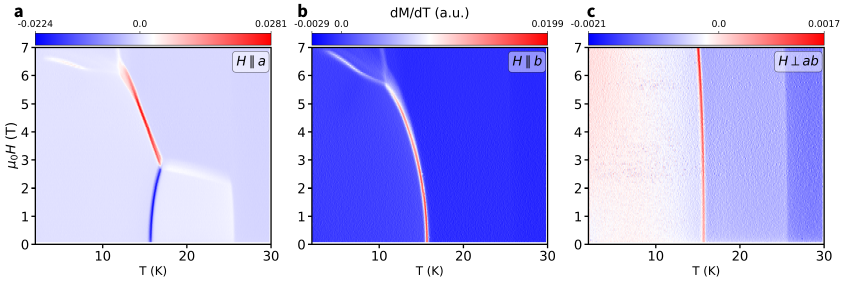}
\caption{\textbf{dM/dT contour maps for \ce{GdNiSn4}.} $dM/dT(T,H)$ values were obtained by differentiating FCW magnetization temperature sweeps at fixed applied fields for three orientations: $H\parallel a$ \textbf{(a)}, $H\parallel b$ \textbf{(b)}, $H\perp ab$ \textbf{(c)}. Extrema in $dM/dT$ tracks the field evolution of the magnetic phase transitions and outline the magnetic phase regimes. The in-plane orientations exhibit multiple field-dependent features, whereas the out-of-plane orientation exhibits two weakly field-dependent transitions.}
\label{GdniSn4_MTPD}
\end{figure}

To map the field evolution of the low-temperature magnetic transitions seen in Fig.~\ref{GdniSn4_MvT_RvT}a, a series of FCW magnetization sweeps were collected at fixed applied fields for two in-plane orientations~($H\parallel a$ and $H\parallel b$) and one out-of-plane orientation~($H\perp ab$). The temperature-dependent derivative, $dM/dT$, was calculated by numerical differentiation of the FCW sweeps and assembled into contour plots of $dM/dT(T,H)$ (Fig.~\ref{GdniSn4_MTPD}). The resulting extrema~(blue and red ridges) tracks the field-evolution of the magnetic phase transitions and roughly outline the magnetic phase regimes for each orientation. 

For $H\perp ab$, the field evolution of both transitions is weakly field-dependent. In contrast, the two in-plane orientations exhibit multiple field-induced transitions ($\approx$ 5-7 T), with $H\parallel a$ additionally showing an extended phase regime between roughly 1-3~T and 15--25~K. The anisotropic magnetic behavior and complex magnetic response for the two in-plane orientations are suggestive of nontrivial magnetic ordering. We have since resolved the magnetic ground state of \ce{GdNiSn4} as a collinear spin density wave state~\cite{GdNiSn4_SDW}, confirming that the new structure type is not merely a crystallographic curiosity but a platform for unconventional magnetism.

\section*{Discussion}
The central result of this work is not that a generative model missed one structure. It is that an experimentally accessible, thermodynamically stable structure type built from known structural motifs could not be recovered from these generative models. The monoclinic \ce{GdNiSn4} and \ce{LuNiSn4} structures are difficult to predict not because their building blocks are unfamiliar, but because the correct structure requires a large unit-cell intergrowth, a specific Wyckoff template, and a packing-compatible slab offset and sequence. The historical \ce{LuNiSn4} assignment also shows the corresponding data problem. When an averaged or incorrect structural model gets entered into structural databases, downstream prediction workflows can inherit and amplify that error.   

What gives us optimism is that the chemistry behind \ce{GdNiSn4} is not unknown. We did not need a generative model to anticipate the structure; we needed the reasoning chemists use routinely. The key was to recognize that the \ce{GdSn2} and \ce{NiSn2}-derived slabs form a stable structural complement. The corrugated \ce{GdSn2} slab provides the chemical-pressure environment needed to accommodate a high-pressure-like \ce{NiSn2} motif at ambient pressure, while the observed lateral offset between slabs further relieves steric frustration and preserves the Sn--Sn bonding motif. Every step of that argument is explicit, and explicit reasoning can be directly translated into machine accessible descriptors. The lesson is therefore constructive. Current generative models lack the chemical principles (descriptors) needed to push its ability to predict materials in such a way that is biased towards an out-of-distribution structure type. We should give it the ability to reason over chemical principles, including electron counting, packing frustration, the covalent Sn-Sn bonding revealed here by ELF, and, most directly, the stacking and intergrowth of known structural units. A model that can propose ``take these two slabs and stack them with this registry'' would generate \ce{GdNiSn4}, because \ce{GdNiSn4} is exactly that. Recombining whole structures is already possible in genetic-algorithm searches~\cite{Glass2006USPEXEvolutionaryPrediction, Trimarchi2007GlobalConstraints}, but the missing capability is chemical: knowing which motifs are compatible, and in what way, before assembling them.

This reframes the relationship between AI and exploratory synthesis as a two-way exchange rather than a contest. AI can dramatically widen and accelerate the search, while experiment supplies the rigorously characterized structures that expand the vocabulary models learn from, including the negative results~\cite{Raccuglia2016Machine-learning-assistedExperiments} and corrected assignments that databases rarely record. New structure types are the most valuable currency in that exchange, precisely because they are the hardest to predict. \ce{GdNiSn4} contributes one such example, together with a worked chemical rationale that could be turned into a model constraint and magnetic properties interesting enough to study in its own right.

We expect that the next generation of materials-prediction tools, combining the speed of generative sampling with the chemical reasoning that solid-state chemists apply intuitively, will predict structures like this one. Showing clearly where current models struggle, and why, is a necessary step toward building the models that will not.

\clearpage 
\bibliography{Reference} 
\bibliographystyle{sciencemag}
\section*{Acknowledgments}
The authors thank Daniel C. Fredrickson, Sally T. Hoang, Austin M. Ferrenti, K. Cooper Stuntz, Joshua P. Wakefield, Kang Zhang, Rina N. Helt, and Michael O' Connor for their assistance with data analysis and for their useful discussions. 
\paragraph*{Funding:}
This work was supported by the Air Force Office for Scientific Research (AFOSR) under grant number FA9550-23-1-0635 (awarded to L.M.S.), the Princeton Center for Complex Materials, a National Science Foundation (NSF)-MRSEC program (DMR-2011750) (L.M.S. and B.A.B) and the National Science Foundation through the AI Research Institutes program Award No.~DMR-2433348 (F.K., L.M.S. and B.A.B.). Equipment used for this work was purchased using funds from the Gordon and Betty Moore Foundation’s EPIQS initiative through Grants GBMF9064, The Packard Foundation, the Princeton Catalysis Initiative (PCI), and the Arnold and Mabel Beckman Foundation via a BYI grant to L.M.S. The authors acknowledge the use of Princeton’s Imaging and Analysis Center, which is partially supported by the Princeton Center for Complex Materials, a National Science Foundation (NSF)-MRSEC program (DMR- 2011750). S.B.L. and C.J.P. were supported by the National Science Foundation Graduate Research Fellowship Program under grant number DGE-2039656. X.Z. was supported by the National Science Foundation Graduate Research Fellowship Program under grant number DGE-2146755. G.S. was supported by the Arnold and Mabel Beckman Foundation via an AOB postdoctoral fellowship (https://doi.org/10.13039/100000997). B.A.B. was supported by the Gordon and Betty Moore Foundation through Grant No. GBMF8685 towards the Princeton theory program, the Gordon and Betty Moore Foundation’s EPiQS Initiative (Grant No. GBMF11070), the Global Collaborative Network Grant at Princeton University, the Simons Investigator Grant No. 404513, the NSF-MERSEC (Grant No. MERSEC DMR 2011750), the Simons Collaboration on New Frontiers in Superconductivity (Grant No. SFI-MPS-NFS-00006741-01), Princeton Catalysis Initiative (PCI), the Schmidt Foundation at the Princeton University and the National Science Foundation through the AI Research Institutes program Award No. DMR-2433348. Y.J. was supported by the European Research Council (ERC) under the European Union’s Horizon 2020 research and innovation program (Grant Agreement No.~101020833), as well as by the IKUR Strategy under the collaboration agreement between Ikerbasque Foundation and DIPC on behalf of the Department of Education of the Basque Government. H.P. was supported by the Ministry for Digital Transformation and of Civil Service of the Spanish Government through the QUANTUM ENIA project call - Quantum Spain project, the European Union through the Recovery, Transformation and Resilience Plan - NextGenerationEU within the framework of the Digital Spain 2026 Agenda, and Gipuzkoa Quantum 2025 grant (ref: 2025-QUAN-000009-01).

\paragraph*{Author contributions:}
X.Z. and L.M.S. conceived and designed the study. X.Z. and E.W. synthesized the crystals. X.Z., S.B.L., and E.W. performed SCXRD. X.Z. and E.W. performed SEM measurements. X.Z., S.C., C.J.P., and J.M. performed physical property characterization measurements of \ce{GdNiSn4} F.K., E.W., H.P., Y.J., and B.A.B performed DFT calculations. D.W. and V.K. performed the PDA calculations and analysis. H.P., Y.J., and B.A.B. performed AI calculations. C.C.T., S.S., and S.D.W. assisted with physical property data analysis. X.Z., S.B.L., and L.M.S. wrote the manuscript with input from all authors.

\paragraph*{Competing interests:}
There are no competing interests to declare.

\paragraph*{Data, code and materials availability:}
Crystallographic data have been deposited at The Cambridge Crystallographic Data Centre (CCDC). All other data needed to evaluate the conclusions in the paper are present in the paper or the Supplementary Materials.


\subsection*{Supplementary materials}
Materials and Methods\\
Supplementary Text\\
Figures S1 to S11\\
Tables S1 to S8\\
Equations S1 to S16\\

\newpage

\renewcommand{\thefigure}{S\arabic{figure}}
\renewcommand{\thetable}{S\arabic{table}}
\renewcommand{\theequation}{S\arabic{equation}}
\renewcommand{\thepage}{S\arabic{page}}
\setcounter{figure}{0}
\setcounter{table}{0}
\setcounter{equation}{0}
\setcounter{page}{1} 
\setcounter{section}{0}
\setcounter{subsection}{0}
\setcounter{subsubsection}{0}

\renewcommand{\thesection}{S\arabic{section}}
\renewcommand{\thesubsection}{\thesection.\arabic{subsection}}
\renewcommand{\thesubsubsection}{\thesubsection.\arabic{subsubsection}}


\begin{center}
\section*{Supplementary Materials for\\ \scititle}

Xin Zhang,$^{1\dagger}$ Scott B. Lee,$^{1\dagger}$ Sudipta Chatterjee,$^{1\dagger}$ Hanqi Pi,$^{2}$ Yi Jiang,$^{2}$\\
Fatmag\"ul Katmer,$^{1}$ Emily G. Ward,$^{1}$ Daniel E. Widdowson,$^{3}$ Charles C. Tam,$^{4}$\\
Sarah Schwarz,$^{4}$ Connor J. Pollak,$^{1}$ Jaime M. Moya,$^{1}$ Grigorii Skorupskii,$^{1}$\\
Vitaliy A. Kurlin,$^{3}$ Stephen D. Wilson,$^{4}$ \\ 
B. Andrei Bernevig,$^{5,2,6\ast}$, Leslie M. Schoop$^{1\ast}$\\
\small{$^{1}$Department of Chemistry, Princeton University, Princeton, New Jersey 08544, USA}\\
\small{$^{2}$Donostia International Physics Center (DIPC), 20018 San Sebasti\'an, Spain}\\
\small{$^{3}$Materials Innovation Factory, University of Liverpool, Liverpool L7 3NY, UK}\\
\small{$^{4}$Materials Department, University of California, Santa Barbara, California 93106, USA}\\
\small{$^{5}$Department of Physics, Princeton University, Princeton, New Jersey 08544, USA}\\
\small{$^{6}$IKERBASQUE, Basque Foundation for Science, Bilbao, Spain}\\
	\small$^\ast$Corresponding author. Email: lschoop@princeton.edu, bernevig@princeton.edu\and
    \\
	\small$^\dagger$These authors contributed equally to this work.

\end{center}

\subsubsection*{This PDF file includes:}
Materials and Methods\\
Supplementary Text\\
Figures S1 to S11\\
Tables S1 to S8\\
Equations S1 to S16\\
\newpage
\section{Materials and Methods}

\subsection{\ce{GdNiSn4} and \ce{LuNiSn4} Synthesis}
Single crystals of \ce{GdNiSn4} were synthesized using the self-flux method\cite{Canfield01061992}. Elemental constituents of gadolinium (Gd) pieces (99.9\%, Thermo Fisher Scientific), nickel (Ni) powder (99.9\%, 3-7 $\mu$m, Thermo Fisher Scientific), and tin (Sn) shot (99.999\%, 3 mm, Beantown Chemical) were loaded together in a fused silica tube of 14 (inner diameter) by 16 (outer diameter) mm in a 1:1.25:297 ratio, where fused quartz wool was utilized as a frit. The tube (loaded with 30 grams of starting material) was then evacuated, purged with Ar three times, and flame-sealed under a 70 mTorr Ar backfill. The tubes are placed upright in a box furnace where they are heated to 1000 \degree{C} at 50 \degree{C/hr} for 24 hours, cooled to 700 \degree{C} at 8 \degree{C/hr}, and then cooled to 300 \degree{C} at 1 \degree{C/hr} before being decanted by inverting and centrifuging the tube. The crystals typically grow as shiny rectangular plates, with the largest face being the $ab$-plane, and they preferentially grow along the $b$-axis (Fig.~\ref{GdniSn4_Crystals}). The crystals are air-stable and retain their physical properties after prolonged exposure to air for several months. 

Single crystals of \ce{LuNiSn4} were synthesized using the self-flux method\cite{Canfield01061992}. Elemental constituents of lutetium (Lu) pieces (99.9\%, Thermo Fisher Scientific), nickel (Ni) powder (99.9\%, 3-7 $\mu$m, Thermo Fisher Scientific), and tin (Sn) shot (99.999\%, 3 mm, Beantown Chemical) were loaded together in a fused silica tube of 14 (inner diameter) by 16 (outer diameter) mm in a 1.25:6:235 ratio, where fused quartz wool was utilized as a frit. The tube (loaded with 30 grams of starting material) was then evacuated, purged with Ar three times, and flame-sealed under a 70 mTorr Ar backfill. The tubes are placed upright in a box furnace where they are heated to 900 \degree{C} at 50 \degree{C/hr} for 24 hours, cooled to 700 \degree{C} at 40 \degree{C/hr}, and then cooled to 430 \degree{C} at 1 \degree{C/hr} before being decanted by inverting and centrifuging the tube. The crystals are air-stable and retain their physical properties after prolonged exposure to air for several months.

\subsection{Single-Crystal X-ray Diffraction}
Single-crystal X-ray diffraction (SCXRD) measurements were performed at 297.6~K (room temperature) using a Rigaku XtaLAB Synergy-S diffractometer equipped with a Mo$\,K_{\alpha}$ ($\lambda=0.71073$~\AA), micro-focus sealed X-ray source, mirror-monochromated optics, and a HyPix-Arc 150 hybrid pixel array detector. Run list generation, frame integration, and data reduction were performed in CrysAlisPro (Version 1.171.43.90). Absorption corrections were applied using a numerical absorption correction based on Gaussian integration over a multifaceted crystal model, together with an empirical absorption correction using spherical harmonics through the SCALE3 ABSPACK scaling algorithm. Full-matrix least-squares refinement on $F^2$ was carried out in JANA2020\cite{Petricek_2023}. The \ce{LuNiSn4} refinement was carried out using the same monoclinic structural model identified for \ce{GdNiSn4}, with details of the refinement and comparison to the previously reported orthorhombic model provided in section~S1 and section~S4 of the SI. 

\subsection{Heat Capacity}
Heat capacity measurements were performed using a Quantum Design 9T Dynacool Physical Properties Measurement System (PPMS) utilizing the semi-adiabatic relaxation method with a 1\% temperature rise measured over 3 time constants\cite{Hwang1997MeasurementCalorimeter}. Heat capacity at each temperature was measured three times through the semi-adiabatic relaxation method and averaged. Samples were mounted on a heat capacity puck using Apiezon N Grease. 

\subsection{Electrical Transport}
For electrical transport measurements, \ce{GdNiSn4} single crystals were cut into a bar-shaped geometry. Longitudinal resistivity was measured using a standard collinear four-probe configuration (Fig.~\ref{GdniSn4_ETO_Device}). Samples were attached to a 0.332 mm thick sapphire substrate (EPI polished one side, $c$-axis, HEMCOR single crystal, Thermo Fisher Scientific) via GE varnish (GE 7031, CMR-Direct). Electrical contacts were prepared using conducting silver paint (DuPont 4922N) and gold wires (25 $\mu$m diameter, 99.9\%, Thermo Fisher Scientific). All transport data were acquired using a Quantum Design 14T Dynacool Physical Properties Measurement System (PPMS) with the Electrical Transport Option (ETO). Residual resistivity ratios (RRR) were calculated as $\mathrm{RRR}=\rho_{300\,\mathrm{K}}/\rho_{4\,\mathrm{K}}$, using 4 K as the low-temperature reference to avoid spurious superconducting contributions from residual Sn flux. All temperature-dependent resistivity data were collected at a ramp rate of 2 Kelvin per minute. The crystallographic axes for the \ce{GdNiSn4} crystals were determined through a combination of crystal morphology and SCXRD. 

\subsection{Magnetic Characterization}
Temperature and field-dependent magnetic susceptibility measurements were performed in a Quantum Design Magnetic Property Measurement System (MPMS3) equipped with the VSM-SQUID option. Samples were mounted onto quartz paddle sample holders using GE varnish (GE 7031, CMR-Direct). The crystallographic axes for the \ce{GdNiSn4} crystals were determined by a combination of crystal morphology and SCXRD. To minimize the individual remnant fields in the SQUID magnet, applied magnetic fields were oscillated to 0 T, and the fields over 2 T were decreased to 2 T linearly and then oscillated to 0 T (Quantum Design, Application Note 1500-021).

\subsection{Training details for MatterGen}
The \textit{MatterGen}~\cite{Zeni2025ADesign} model used for benchmarking was trained on the MPTS-52 dataset using the official CSP training configuration. 
Training was performed in PyTorch Lightning on a single NVIDIA A100 GPU using the Adam optimizer. The learning rate is initialized at \(1\times10^{-4}\) and decayed using \texttt{ReduceLROnPlateau} scheduler (factor 0.6, patience 100, minimum learning rate \(1\times10^{-6}\)). The training loss includes position and cell terms with weights 0.1 (position) and 1.0 (cell), respectively.
We trained for 1000 epochs and retained the checkpoint with the lowest validation loss (epoch 964, validation loss 0.727). The validation loss had plateaued over the final few hundred epochs, improving by only about 2\% beyond epoch 700, indicating that the model had converged. The trained model achieves a 26.9\% SUN (stable, unique, and novel) rate over 5{,}000 generated structures relaxed with the Orb force field~\cite{neumann2024orb}, where stability and novelty are assessed against a reference dataset composed of the Materials Project and Alexandria databases ($3{,}997{,}799$ structures), using a convex-hull threshold of $0.1$~eV/atom. This rate is comparable to the 38.57\% and 22.27\% SUN rates reported on the Alex-MP-20 and MP-20 datasets, respectively.

\subsection{First Principles Calculations}
Density functional theory (DFT) calculations were performed as implemented in the Vienna ab initio simulation package (VASP) 6.4.2 \cite{Kresse1993AbMetals, Kresse1994AbGermanium, Kresse1996EfficiencySetb, Kresse1996EfficientSet}. The Perdew–Burke–Ernzerhof (PBE) generalized gradient approximation (GGA)\cite{Perdew1996GeneralizedSimplec} functional was used for exchange-correlation with the recommended potpaw-PBE.64 projector-augmented wave (PAW) pseudopotentials provided by VASP\cite{Blochl1994ProjectorMethod, Kresse_ultrasoft_1999}. Initial calculations used a converged (criterion of $<10^{-4}$ eV) plane-wave energy cut-off of 650 eV, a converged (criterion of $<10^{-3}$ eV) $\Gamma$-centered $9\times9\times9$ Monkhorst-Pack $\mathbf{k}$-point grid \cite{Monkhorst_kpoints_1976}, and $f$-electrons were set in the core. Experimentally determined structures were relaxed with the convergence criteria of $<10^{-5}$ eV energy difference between steps and used in all subsequent calculations. Energy calculations varied the exchange correlation functional to be the meta-GGA Regularized Strongly Constrained and Appropriately Normed (RSCAN) semilocal \cite{Bartok2019_RSCAN} functional, the $\mathbf{k}$-point grid to $15\times15\times15$, the presence of spin-orbit coupling, and the $f$-electrons were moved from the core to the valence shell. Electron Localization Function (ELF) plots were calculated by VASP after calculations without spin-orbit coupling (SOC) \cite{ELF1_silvi_1994, ELF2_becke_1990, ELF3_burdett_1998, Savin1997ELF:Function, Savin1992ElectronStructure}. Band structures and density of states (DOS) plots were calculated with the inclusion of spin-orbit coupling. The projected density of states (PDOS) and band structure visualizations were plotted using the sumo package \cite{SUMOPackage}, and projected band structures were plotted using the Modeling and Crystallographic Utilities (MCU) package \cite{MCUPackage}. Ferromagnetic and antiferromagnetic DFT+U calculations were performed utilizing the same convergence criterion with the Dudarev formalism \cite{dudarev_dftu_1998, anisimov_band_1991, rohrbach_electronic_2003}. The $U' = U - J$ parameter was taken as 6 eV in accordance with previous literature \cite{petersen_gdu_2006}, although results from $U'$ = 0 -- 8 eV variants are provided in Section 9 of the SI, and demonstrate the same trend, with an initial magnetic moment of 6.8 $\mu_B$ on each \ce{Gd} site. For antiferromagnetic calculations, the $\Gamma$-centered $9\times9\times9$ Monkhorst-Pack $\mathbf{k}$-point grid was reduced to $5\times5\times5$ to accommodate the much larger unit cell size \cite{Monkhorst_kpoints_1976}. Further details are provided in Supplemental Section 9. 
For the evaluation of energies above the convex hull, we follow the Materials Project~\cite{jain2013commentary} \texttt{MPRelaxSet} convention, using the VASP PAW-PBE \texttt{PBE\_52}~POTCAR set, with \textsc{pymatgen}~\cite{ong2013python} using Materials-Project-compatible phase-diagram entries and corrections. 

\subsection{PDA Calculation Details}
we searched both the ICSD and the Materials Project for closely related structures using a generically complete invariant descriptor PDA (Pointwise Deviation from Asymptotic), which distinguishes all non-duplicate structures in major databases of experimental materials \cite{widdowson2022resolving,widdowson2025geographic, widdowson2026pointwise}. The PDA distance between two periodic structures measures the maximum difference between the distance vectors to the $k$ atomic neighbors of optimally matched atoms. If every atom in a given structure is perturbed up to a small $\epsilon>0$, the PDA distance between the original structure and its perturbation is at most $2\epsilon$~(Theorem~4.2)~\cite{widdowson2026pointwise}. Any PDA distance $d>0$ means that matching two structures exactly requires shifting some atoms by at least $d/2$ in Angstroms (\AA).T he full process for finding the closest matches of a structure by PDA in any database (e.g. ICSD) is as follows (experiments were ran on a typical desktop computer, AMD Ryzen 5 5600X, 32~GB RAM). We first compute PDA for all items in the database. In this desktop setup, 170,513 non-disordered ICSD structures can be processed in approximately 35 minutes (12.19~ms per CIF). Before comparing by PDA, we make use of the PDA's column-wise average, which is a vector called average deviation from asymptotic (ADA). Because of a theorem relating PDA and ADA distances, we can search using the simpler ADA first to get a candidate list of matches, which uses a KD-tree (binary search). The tree is first built on the target database, then queried with the desired structure(s). Once PDA and ADA are computed, they do not need to be recomputed for later comparisons, and the KD-tree can also be pre-computed.


\section{Supplementary Text}

\subsection{\ce{GdNiSn4} and \ce{LuNiSn4} SCXRD} \label{SCXRD}
Table S1 summarizes the data collection and refinement parameters for \ce{GdNiSn4}. The SCXRD refinement is in good agreement with the experimental data. Table S2 lists the atomic positions of \ce{GdNiSn4} and their corresponding displacement parameters. Table S3 contains the anisotropic atomic displacement parameters used to calculate $U_{eq}$. Table S4 summarizes the data collection and refinement parameters for \ce{LuNiSn4}. The SCXRD refinement is in good agreement with the experimental data. Table S5 lists the atomic positions of \ce{LuNiSn4} and their corresponding displacement parameters. Table S6 contains the anisotropic atomic displacement parameters used to calculate $U_{eq}$ for \ce{LuNiSn4}. 

\begin{center}
\begin{table} [H]
\renewcommand{\tablename}{Table S}
\caption{\textbf{Crystallographic Data} for the refinement of \ce{GdNiSn4}.}
 \centering
 \begin{tabular}{l c}
\hline
Refined Composition & \ce{GdNiSn4} \\ 
\hline 
Crystal Dimension (mm) & 0.026 $\times$ 0.079 $\times$ 0.125  \\ 
Radiation source, $\lambda$ (\AA) & X-ray, 0.71073 \\
Absorption Correction & multi-scan \\
Data Collection Temperature (K)  & 297.64(15) \\
Space Group & C2/m (12) \\
$a$ (\AA) & 8.74042(11) \\  
$b$ (\AA) & 8.82366(8) \\
$c$ (\AA) & 14.33415(14) \\ 
$\alpha$ (\degree) & 90  \\
$\beta$ (\degree) & 98.7710(9) \\
$\gamma$ (\degree) & 90 \\
Cell Volume (\AA$^{3}$)  &  1092.56(2) \\ 
Absorption Coefficient (mm$^{-1}$) & 33.177 \\ 
$\theta_{min}$ , $\theta_{max}$ (deg) & 2.88, 40.68 \\ 
Refinement Method & F$^{2}$   \\
R$_{int}$($I > 3\sigma$)  & 0.0854\\
Deposition Number & 2535295 \\
\hline 
\multicolumn{2}{c}{\textit{Overall Refinement}} \\
\hline 
Total Reflections ($I > 3\sigma$, all)  & 31607, 60790   \\ 
Unique Reflections ($I > 3\sigma$, all)  & 2521, 3645  \\ 
Number of Parameters                         & 75  \\ 
R(all), R$_{w}$(all)                         &   0.0352, 0.0575 \\
R($I > 3\sigma$), R$_{w}$($I > 3\sigma$)     &   0.0226, 0.0501 \\ 
S($I > 3\sigma$), S(all)                     & 1.1275, 1.0727 \\ 
$\Delta\rho_{max}$ , $\Delta\rho_{min}$ (e \AA$^{-3}$) & 1.76, -1.81 \\ 
\label{R_114}
\end{tabular}
\end{table}
\end{center}

\begin{center}
\begin{table} [H]
\renewcommand{\tablename}{Table S}
\caption{\textbf{Atomic positions} for \ce{GdNiSn4}.}
 \centering
 \begin{tabular}{c c c c c c c}
\hline
Site  & Wyckoff Position  & $x$ &  $y$  & $z$ &  Occupancy &  U$_{eq}$ \\ 
\hline
  Gd1 & 4i   & 0.40434(7)   & 0.5         & 0.118996(19) &  1  & 0.01009(8)  \\ 
  Gd2 & 4i   & 0.40457(7)   & 1           & 0.118846(19) &  1  & 0.01010(8)  \\ 
  Sn1 & 8j   & 0.64554(12)  & 0.74981(7)  & 0.06994(4)   &  1  & 0.01152(9)  \\ 
  Sn2 & 4i   & 0.69822(12)  & 1           & 0.29088(3)   &  1  & 0.01170(11) \\ 
  Sn3 & 8j   & 0.44736(11)  & 0.75062(3)  & 0.289984(15) &  1  & 0.01156(9)  \\ 
  Sn4 & 4i   & 0.69729(12)  & 0.5         & 0.29085(3)   &  1  & 0.01163(11) \\ 
  Sn5 & 4i   & 0.5          & 0.66224(4)  & 0.5          &  1  & 0.01020(9)  \\ 
  Sn6 & 4h   & 0.66276(4)   & 1           & 0.50002(5)   &  1  & 0.01025(9)  \\
  Ni1 & 8j   & 0.7263(3)    & 0.75007(7)  & 0.40540(3)   &  1  & 0.01138(15) \\
\hline
\label{APP}
\end{tabular}
\end{table}
\end{center}

\begin{center}
\begin{table} [H]
\renewcommand{\tablename}{Table S}
\caption{\textbf{Components of the anisotropic ADP (U$_{ani}$) parameters} for \ce{GdNiSn4}.}
 \centering
 \begin{tabular}{c c c c c c c}
\hline
Atom & U$_{11}$ & U$_{22}$ & U$_{33}$ & U$_{12}$ & U$_{13}$ & U$_{23}$ \\ 
\hline
Gd1 & 0.01037(15)   & 0.01048(10) & 0.00947(10) & 0                 & 0.0017(3)  & 0  \\ 
Gd2 & 0.01016(15)   & 0.01062(10) & 0.00953(10) & 0                 & 0.0015(3)  & 0   \\
Sn1 & 0.00978(18)   & 0.01170(10) &0.01322(10)  &0.00010(10)        &0.0023(4)   &0.00006(9) \\
Sn2 & 0.0121(2)     &0.01260(15)  &0.01064(15)  &0                  &0.0024(4)   &0   \\
Sn3 & 0.01149(17)   &0.01282(9)   &0.01049(9)   &0.00004(10)        &0.0020(4)   &0.00027(9)   \\ 
Sn4 & 0.0116(2)     &0.01238(15)  &0.01092(15)  &0                  &0.0019(4)   &0  \\
Sn5 & 0.00861(16)   &0.01075(13)  &0.01104(13)  &0                  &0.0008(2)   &0   \\
Sn6 & 0.00974(17)   &0.00978(13)  &0.01102(13)  &0                  &0.0008(2)   &0  \\
Ni1 & 0.0110(3)     &0.01286(17)   &0.01018(16) &\text{-}0.0002(2)  &0.0011(7)   &0.00009(18)   \\
\hline
\label{U_APP}
\end{tabular}
\end{table}
\end{center}

\begin{center}
\begin{table} [H]
\renewcommand{\tablename}{Table S}
\caption{\textbf{Crystallographic Data} for the refinement of \ce{LuNiSn4}.}
 \centering
 \begin{tabular}{l c}
\hline
Refined Composition & \ce{LuNiSn4} \\ 
\hline 
Crystal Dimension (mm) & 0.091 $\times$ 0.075 $\times$ 0.029  \\ 
Radiation source, $\lambda$ (\AA) & X-ray, 0.71073 \\
Absorption Correction & multi-scan \\
Data Collection Temperature (K)  & 293.15(10) \\
Space Group & C2/m (12) \\
$a$ (\AA) & 8.71521(7)  \\
$b$ (\AA) & 8.77764(7)  \\
$c$ (\AA) & 14.04356(12)\\ 
$\alpha$ (\degree) & 90  \\
$\beta$ (\degree) & 98.9257(7)\\
$\gamma$ (\degree) & 90 \\
Cell Volume (\AA$^{3}$)  &  1061.308(15) \\ 
Absorption Coefficient (mm$^{-1}$) & 40.26 \\ 
$\theta_{min}$ , $\theta_{max}$ (deg) & 2.94, 44.95 \\ 
Refinement Method & F$^{2}$   \\
R$_{int}$($I > 3\sigma$)  & 0.0448\\
Deposition Number & TBD \\
\hline 
\multicolumn{2}{c}{\textit{Overall Refinement}} \\
\hline 
Total Reflections ($I > 3\sigma$, all)   & 37311, 60962   \\ 
Unique Reflections ($I > 3\sigma$, all)  & 3992, 4561  \\ 
Number of Parameters                         & 65  \\ 
R(all), R$_{w}$(all)                         &   0.0240, 0.0795 \\
R($I > 3\sigma$), R$_{w}$($I > 3\sigma$)     &   0.0202, 0.0786 \\
S($I > 3\sigma$), S(all)                     &   2.6025, 2.4605 \\
$\Delta\rho_{max}$ , $\Delta\rho_{min}$ (e \AA$^{-3}$) & 3.25, -2.47 \\ 
\label{R_114_LuNiSn4}
\end{tabular}
\end{table}
\end{center}

\begin{center}
\begin{table} [H]
\renewcommand{\tablename}{Table S}
\caption{\textbf{Atomic positions} for \ce{LuNiSn4}.}
 \centering
 \begin{tabular}{c c c c c c c}
\hline
Site  & Wyckoff Position  & $x$ &  $y$  & $z$ &  Occupancy &  U$_{eq}$ \\ 
\hline
  Lu1 & 4i   & 0.09642(2)  & 0            & -0.114691(13)&  1  & 0.01062(4)  \\ 
  Lu2 & 4i   & 0.09627(2)  & 0.5          & -0.114428(13)&  1  & 0.01059(4)  \\ 
  Sn1 & 4h   & 0           & 0.16220(3)   & -0.5         &  1  & 0.00970(6)  \\ 
  Sn2 & 4i   & 0.16256(3)  & 0.5          & -0.50003(2)  &  1  & 0.00968(6)  \\ 
  Sn3 & 8j   & 0.35686(3)  & 0.250134(18) & -0.072673(18)&  1  & 0.01109(5)  \\ 
  Sn4 & 4i   & 0.30343(4)  & 0            & -0.28577(2)  &  1  & 0.01113(6)  \\ 
  Sn5 & 4i   & 0.30372(4)  & 0.5          & -0.28578(2)  &  1  & 0.01115(6)  \\ 
  Sn6 & 8j   & 0.05384(3)  & 0.250318(18) & -0.284782(17)&  1  & 0.01111(5)  \\
  Ni1 & 8j   & 0.27425(6)  & 0.25004(4)   & -0.40306(3)  &  1  & 0.01066(9)  \\ 
\hline
\label{APP_LuNisn4}
\end{tabular}
\end{table}
\end{center}

\begin{center}
\begin{table} [H]
\renewcommand{\tablename}{Table S}
\caption{\textbf{Components of the anisotropic ADP (U$_{ani}$) parameters} for \ce{LuNiSn4}.}
 \centering
 \begin{tabular}{c c c c c c c}
\hline
Atom & U$_{11}$ & U$_{22}$ & U$_{33}$ & U$_{12}$ & U$_{13}$ & U$_{23}$ \\ 
\hline
Lu1 & 0.01299(7)    & 0.00967(7)  & 0.00900(8)  & 0                 & 0.00110(5) & 0  \\
Lu2 & 0.01272(7)    & 0.00980(7)  & 0.00907(8)  & 0                 & 0.00111(5) & 0  \\
Sn1 & 0.00989(9)    & 0.00875(9)  & 0.01036(11) & 0                 & 0.00129(8) & 0  \\
Sn2 & 0.01014(9)    & 0.00806(9)  & 0.01074(11) & 0                 & 0.00131(8) & 0  \\
Sn3 & 0.01103(8)    & 0.01045(8)  & 0.01172(9)  & -0.00002(5)       & 0.00152(7) & 0.00006(5)\\ 
Sn4 & 0.01302(10)   & 0.01037(10) & 0.00992(11) & 0                 & 0.00148(8) & 0  \\ 
Sn5 & 0.01280(10)   & 0.01034(10) & 0.01007(12) & 0                 & 0.00097(8) & 0  \\ 
Sn6 & 0.01211(8)    & 0.01135(8)  & 0.00973(9)  & -0.00007(5)       & 0.00124(6) & -0.00017(5)\\
Ni1 & 0.01211(15)   & 0.01073(15) & 0.00898(16) & 0.00005(10)       &0.00115(13) &-0.00008(10)\\
\hline
\label{U_APP_LuNiSn4}
\end{tabular}
\end{table}
\end{center}

\subsection{\ce{GdNiSn4} and \ce{LuNiSn4} Twinning Information}
Twinning percentage was determined for single-crystal X-ray diffraction measurements replicating what is reported elsewhere for \ce{GdNiSn4} \cite{Tam2026}. 

Single-crystal diffraction intensities were obtained via full 3D reciprocal-space integration of the reflection profiles, as implemented in CrysAlisPro. The integrated intensities were exported as an HKL file and imported into Jana2020, where the structure solution and refinement were performed using a twinning model incorporating the twin law (180\textdegree~rotation about $a$) and refinement of the domain volume fractions.

The single-crystal X-ray diffraction structural solution for \ce{LuNiSn4} was refined as a two-component twin along the crystallographic $a$-axis, corresponding to the matrix:

\begin{equation}
T_{M1}= \begin{bmatrix}
\;\;\;\;1\;\;\;\;\;\;\;\;0\;\;\;\;\;\;\; 0\\
\;\;\;\;0\;\; \;\;\;-1\;\;\;\;\; 0\\
-0.5\;\; \;\;\;\;0 \;\;\; -1\\
\end{bmatrix}
\end{equation}

The twinning percentage for the specific solution here was 96\% and not merely a 50\% occupation of two nearly identical sites.

\begin{figure}
    \centering
    \includegraphics[width=1\linewidth]{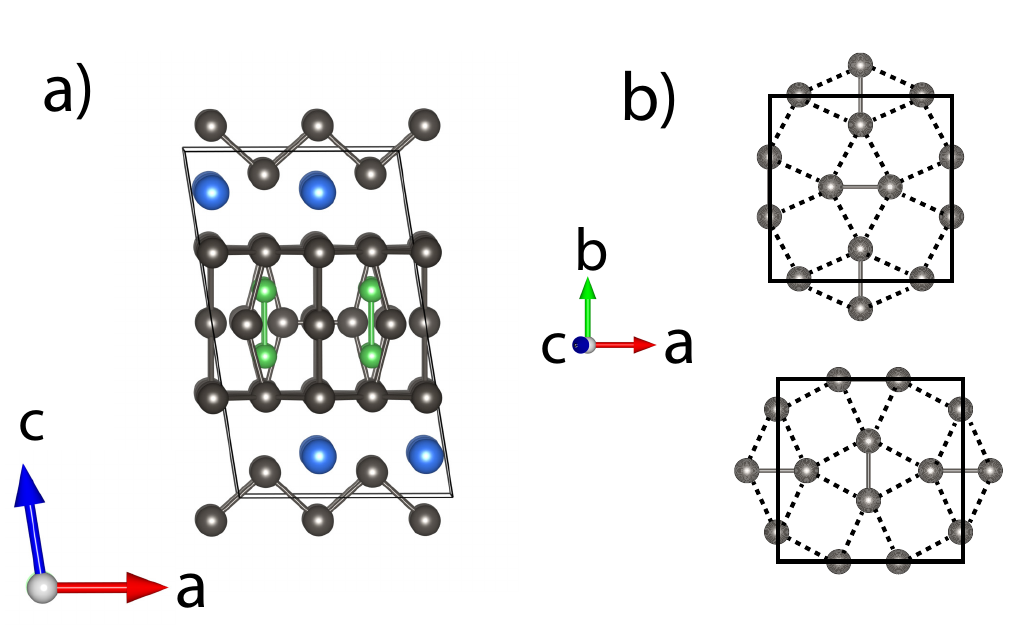}
    \caption{a) structure of monoclinic \ce{RNiSn4} which hosts a $3^{2}434$ net layer in the center ($c$ = 0.5) $a$-$b$ plane. b) This net is susceptible to a in-plane twinning via a 90 degree rotation perpendicular to the plane.}
    \label{monoclinic_figure}
\end{figure}

The main goal of using this twinning matrix was to mathematically model a four-fold rotation perpendicular to the $3^{2}434$ Sn net layer. The pattern of this layer of Sn atoms makes it susceptible to such twinning, as shown in Fig.~\ref{monoclinic_figure}a. However, simply applying the matrix corresponding to a 90$\degree$ rotation along $c$ is insufficient in the monoclinic cell, since the monoclinic angle $\beta$ is not 90$\degree$. Furthermore, a rotation of 90$\degree$ in either $c$ or $c*$ also models the zigzag chains, stretching along the $a$-axis to now be modeled along the $b$-axis. As shown in Fig.~\ref{monoclinic_figure}a, this $3^{2}434$ remains invariant along two-fold rotations along $c*$.

\subsubsection{Cell Transformation}
The transformation from orthorhombic fractional coordinates to monoclinic fractional coordinates is defined by how the monoclinic basis vectors relate to the orthorhombic ones. Since the monoclinic $a$ and $b$-axes are unchanged, and the monoclinic $c$ axis spans one orthorhombic layer (\textit{i.e.}, $\frac{1}{4}$ of the orthorhombic $c'$) while incorporating a shear of $\frac{1}{4}$ along $a$. The corresponding transformation matrix, \textbf{P} is:

\begin{equation}
P= \begin{bmatrix}
1\;\; 0\;\; 0\\
0\;\; 1\;\; 0\\
\frac{1}{4} \;\; 0 \;\; \frac{1}{4}\\
\end{bmatrix}
\end{equation}

While the inverse transformation is: 

\begin{equation}
P^{-1}= \begin{bmatrix}
\;\;\;1\;\;\, 0\;\;\; 0\\
\;\;\,0\;\;\, 1\;\;\; 0\\
-1\;\; 0 \;\;\; 4\\
\end{bmatrix}
\end{equation}

In the orthorhombic aristotype of \ce{LuNiSn4}, a two-fold rotation about $c$ can then be applied, which in orthorhombic fractional coordinates is simply:

\begin{equation}
T_O= \begin{bmatrix}
-1\;\; \;\;\;0\;\;\;\;\;\; 0\\
\;\;\;0\;\;\; -1\;\;\; 0\\
\;\;\;0\;\;\;\;\; 0\;\;\;\;\;\; 1\\
\end{bmatrix}
\end{equation}

Therefore, in order to correctly account for this two-fold rotation perpendicular to the $3^{2}434$ Sn net layer, we can compute:

\begin{equation}
T_{M2}= P \cdot T_O \cdot P^{-1}
\end{equation}

Evaluating this expression yields:

\begin{equation}
T_{M2}= \begin{bmatrix}
\;\;\;\;1\;\;\;\;\;\;\;\;0\;\;\;\;\;\;\; 0\\
\;\;\;\;0\;\; \;\;\;-1\;\;\;\;\; 0\\
-0.5\;\; \;\;\;\;0 \;\;\; -1\\
\end{bmatrix}
\end{equation}

\begin{figure}
    \centering
    \includegraphics[width=1\linewidth]{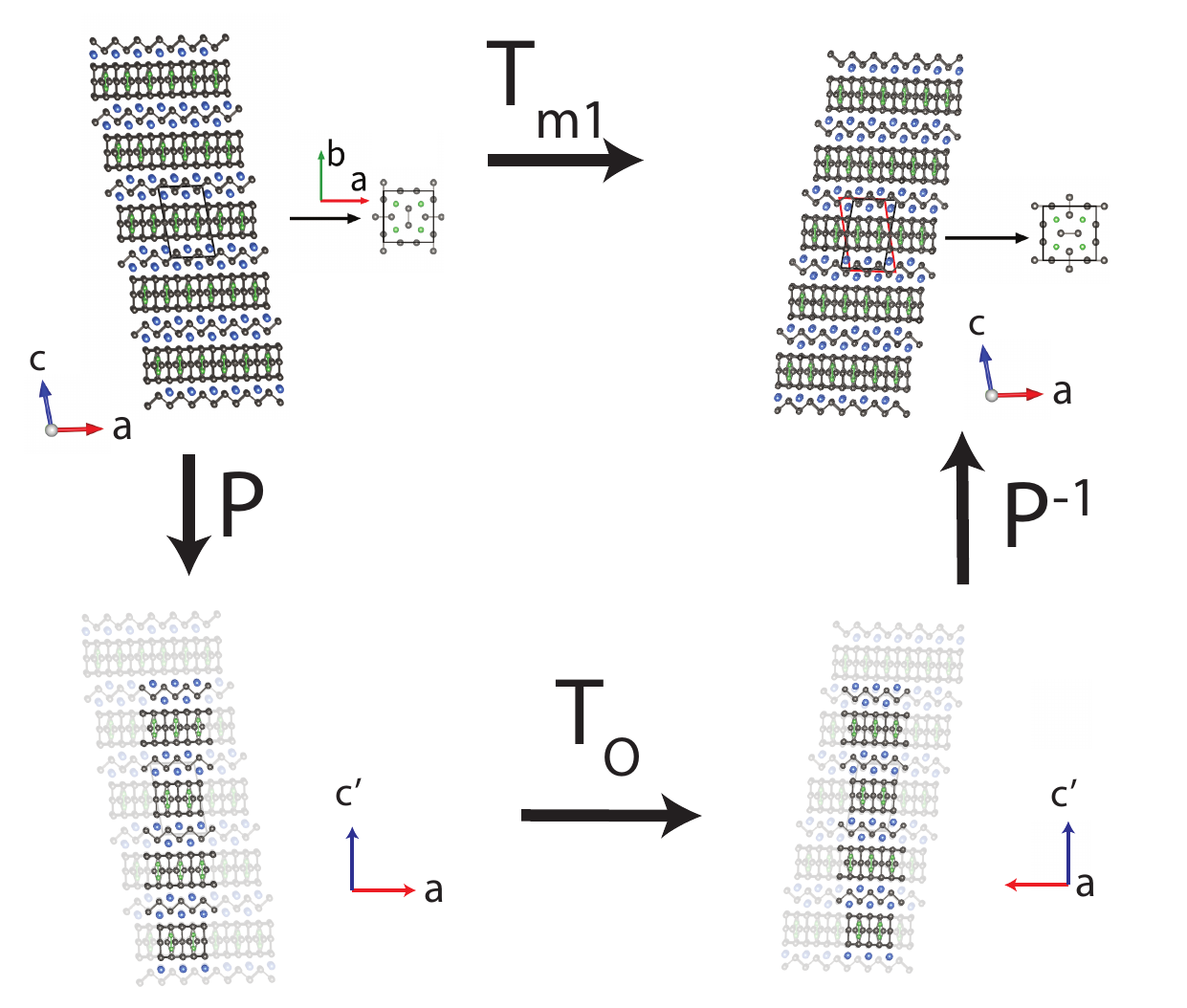}
    \caption{Schematic of the two-fold rotation about the $a$-axis for \ce{RNiSn4}. Starting from the reported monoclinic structure (top left), a transformation of T$_{M1}$ is mathematically equivalent to $P \cdot T_O \cdot P^{-1}$ as defined in the text. Both of these operations result in modeling a twinning of the $3^{2}434$ net (top right).}
    \label{fig:SI_Charles2}
\end{figure}

This matches precisely with the twin matrix modeled in Jana2020 ($T_{M1}$). The off-diagonal element is consistent with the ratio of $-2cos(\beta) \cdot \frac{c}{a}$ evaluated using the refined lattice parameters a = 8.71 \AA, c = 14.04 \AA, and $\beta$ = 98.92 $\degree$. In the monoclinic coordinate system, this off-diagonal term in position arises because $a$ and $c$ are not orthogonal. When modeling a twin rotated 90 $\degree$ about $c$, the $x$-coordinate contributes to the transformed $z$-coordinate through the metric of the cell. Therefore, utilizing the twin matrix related to a two-fold rotation along the crystallographic $a$-axis is mathematically equivalent to the desired four-fold rotation perpendicular to the $3^{2}434$ net, as shown in Fig.~\ref{fig:SI_Charles2}. 

\section{\ce{GdNiSn4} Crystals}
\begin{figure}[H]
\centering
\includegraphics[width=1.00\textwidth]{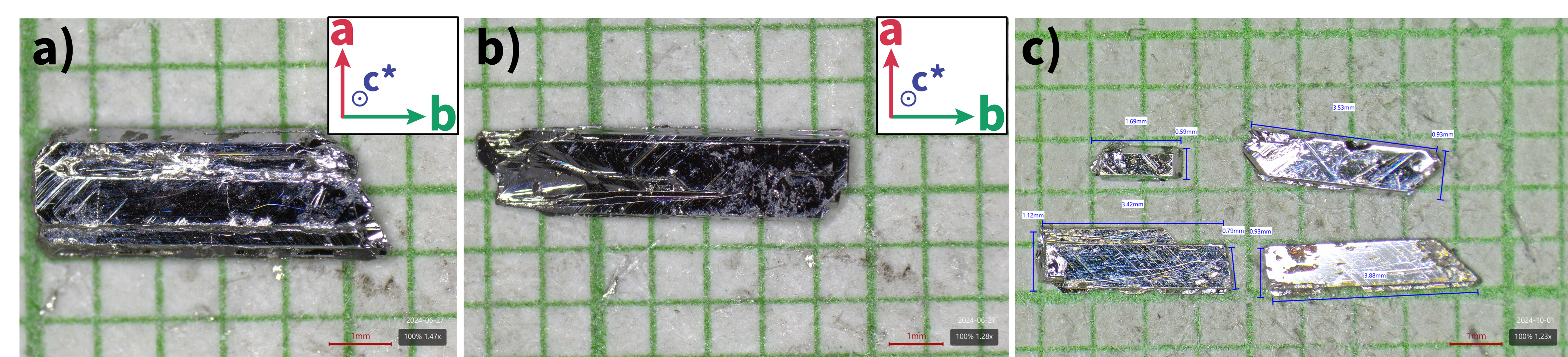}
\caption{\textbf{\ce{GdNiSn4} crystals grown through the self-flux method.} (a,b) \ce{GdNiSn4} single crystals through an $ab$-plane viewpoint. (c) Crystals typically grow as rectangular plates. The largest face is the $ab$-plane, and the crystals grow preferentially along the crystallographic $b$-axis.  }
\label{GdniSn4_Crystals}
\end{figure}

\addcontentsline{toc}{section}{S3: \ce{GdNiSn4} Crystals}
Fig.~\ref{GdniSn4_Crystals} shows various crystals of \ce{GdNiSn4} grown through the self-flux method\cite{Canfield01061992}. The largest face is the $ab$-plane of the crystal, while the crystal grows the longest along the crystallographic $b$-axis. 

\section{SEM-EDS}
\addcontentsline{toc}{section}{S4: SEM-EDS}
Scanning electron microscopy (SEM) was used to visualize samples. SEM images were collected using a Quanta 200 FEG Environmental-SEM in high-vacuum mode at an accelerating voltage of 20~keV. Multiple Energy dispersive X-ray (EDX) spectra were collected to determine the elemental composition of a \ce{GdNiSn4} single crystal. 

\begin{figure}[H]
\centering
\includegraphics[width=1.00\textwidth]{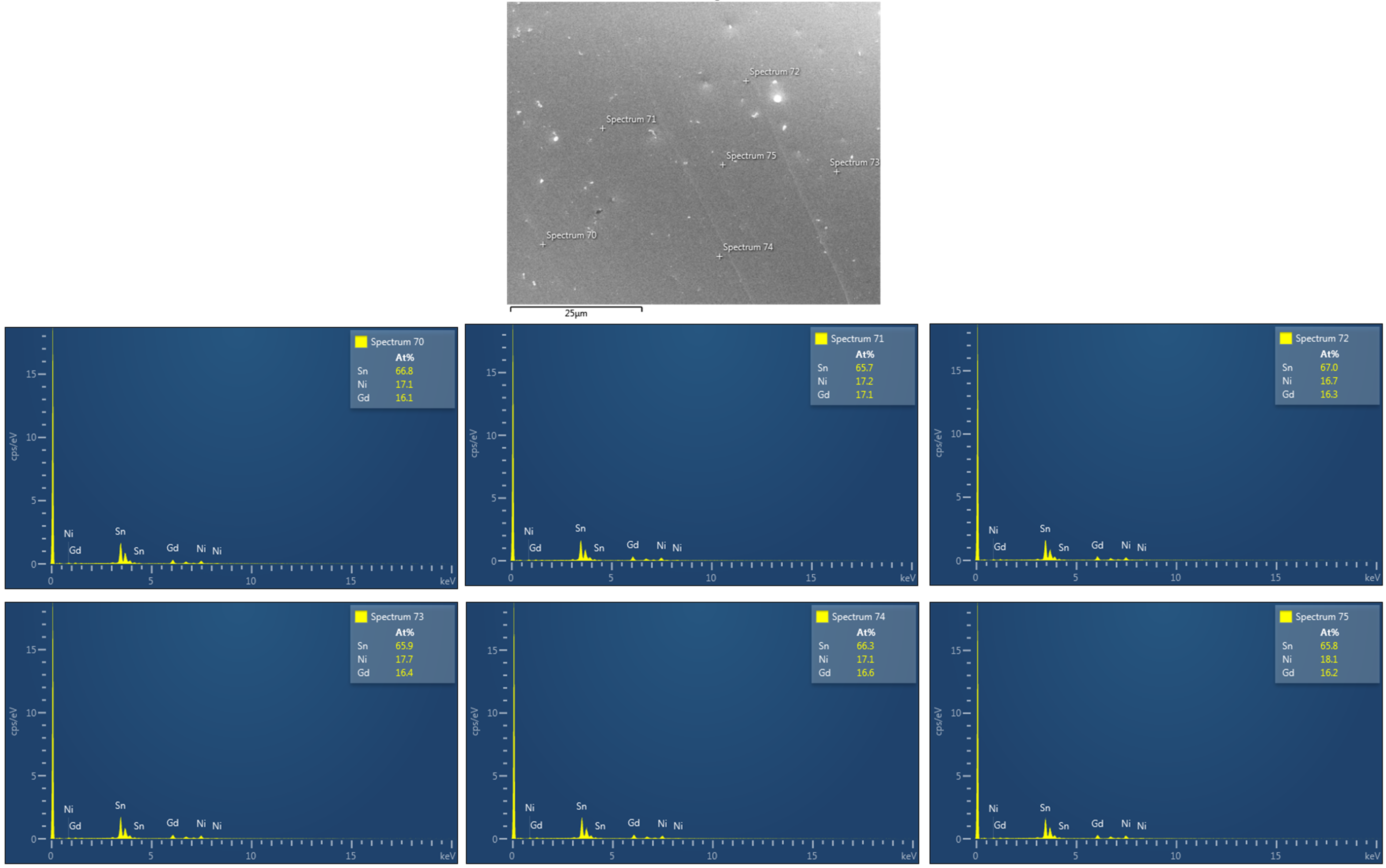}
\caption{\textbf{SEM image of a \ce{GdNiSn4} single crystal and its associated EDX spectra.} All EDX measurements show a nominal compositional ratio that matches that of \ce{GdNiSn4}.}
\label{GdniSn4_SEM}
\end{figure}

From Fig.~\ref{GdniSn4_SEM}, it can be seen that for all six EDX measurements, the elemental composition of the single crystal matches that of \ce{GdNiSn4}.

\begin{figure}[H]
\centering
\includegraphics[width=1.00\textwidth]{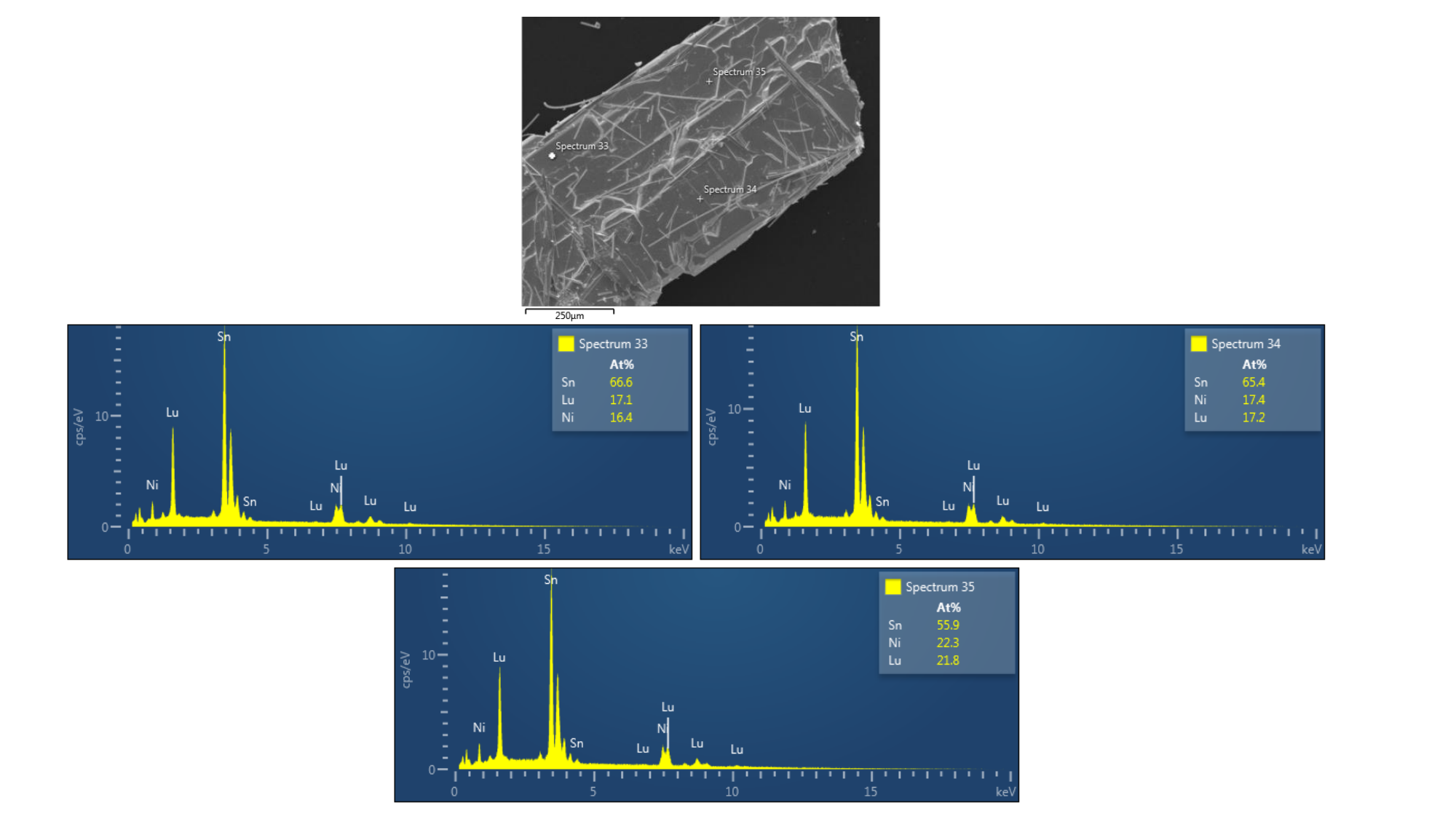}
\caption{\textbf{SEM image of a \ce{LuNiSn4} single crystal and its associated EDX spectra.} All EDX measurements show a nominal compositional ratio that matches that of \ce{GdNiSn4}.}
\label{LuniSn4_SEM}
\end{figure}

From Fig.~\ref{LuniSn4_SEM}, it can be seen that for all six EDX measurements, the elemental composition of the single crystal matches that of \ce{LuNiSn4}.

\section{\ce{GdNiSn4} and \ce{LuNiSn4} comparison}
\addcontentsline{toc}{section}{S5: \ce{GdNiSn4} and \ce{LuNiSn4} comparison}
As stated in the main text, single-crystal X-ray diffraction was previously reported, published in a commensurately modulated supercell with superspace group A$mmm$/($ddd$)($\frac{1}{2}$ $\frac{1}{2}$ $\frac{1}{2}$), and lattice parameters a = 4.3552, b = 4.4039, and c = 22.044 \AA. The reported average structure contains slabs of \ce{PtHg2}-type \ce{NiSn2}, described by $4^{4}$ nets stacking along the crystallographic $c$-axis.

Three observations from the prior structural characterization suggest that \ce{LuNiSn4} likely adopts the newly reported \ce{GdNiSn4} structure type~\cite{Skolozdra2000NewProperties}. The authors initially indexed the data to an orthorhombic cell with a very long $b$-axis ($b=55.48\,$\AA). Subsequent refinement in this unit cell yielded nonphysical results, producing extremely short interatomic Sn--Sn distances on the order of $1.0-1.5\,$\AA\ and partial Sn-site occupancies of 50\%. These issues suggest that the orthorhombic assignment is not an adequate structural description of \ce{LuNiSn4}.

\begin{figure}[H]
\centering
\includegraphics[width=1.00\textwidth]{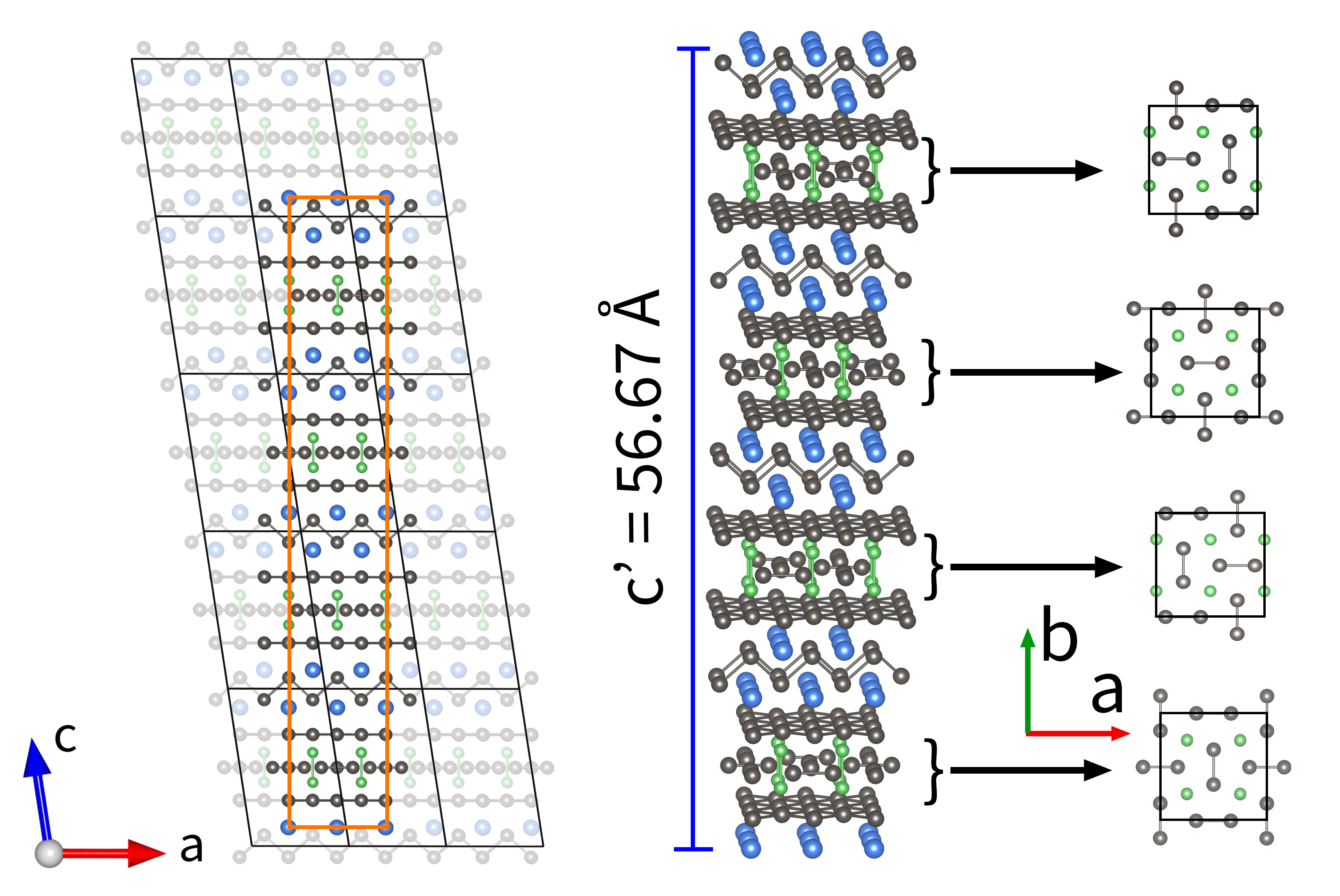}
\caption{\textbf{\ce{GdNiSn4} structure in orthorhombic setting.} (left) a 3$\times$3$\times$5 supercell of \ce{GdNiSn4} in its primitive unit cell. In orange, an orthorhombic unit cell may be drawn with 4 times the volume of the monoclinic cell. (center) Isolates this unit cell, with a new lattice parameter $c$ = 56.67 \AA, very close to what was originally reported for \ce{LuNiSn4}. (right) view of the $ab$ plane for each $3^{2}434$ net along the orthorhombic unit cell. The inequivalence of these layers is likely what leads \ce{LuNiSn4} to be indexed initially to this unit cell. Structure visualized and bond distances calculated through VESTA\cite{VESTA}.}
\label{Structure_Figure}
\end{figure}

In Fig.~\ref{Structure_Figure} we first show a 3$\times$3$\times$5 supercell of our solved \ce{GdNiSn4} structure. In this configuration, an orthorhombic cell with 4 times the volume of our primitive unit cell may be drawn. Examination of the $3^2434$ layer stacking in the orthorhombic cell reveals that each consecutive layer along $c$ is shifted along $\frac{-1}{4}a$ of a unit cell, making this a valid choice as a unit cell. This $3^{2}434$ sublayer stacking, unique to our structure solution, is likely why the initial structure solution for \ce{LuNiSn4} failed while incorporating \ce{PtHg2}-type \ce{NiSn2} units.

This $3^{2}434$ sublayer can also explain the unjustifiably short interatomic distances observed in the \ce{LuNiSn4}-type structure. This $3^2434$ sublayer can be visualized as a $4^{4}$ net where the Sn atoms form dimers. This dimerization causes these layers to become unequal when visualized in the orthorhombic cell (Fig.~\ref{Structure_Figure}). Modeling this layer as a $4^{4}$ Sn square-net would lead to spurious excess electron density being found $\sim$ $1.0-1.5\,$\AA\ away from sites of an ideal $4^{4}$ Sn square-net.

\section{Electron Localization Function (ELF)} \label{ELF_SI}
\addcontentsline{toc}{section}{S6: Electron Localization Function (ELF)}
The electron localization function (ELF) is a metric introduced by Becke and Edgecombe, based on the Pauli exclusion principle, that is used to analyze chemical bonding for both molecular and periodic systems in real space~\cite{ELF1_silvi_1994, ELF2_becke_1990, ELF3_burdett_1998, Savin1997ELF:Function, Savin1992ElectronStructure}. It is commonly used as an indicator for the presence of covalent interactions.

We first define the ratio $\chi$ as

\begin{equation}
\chi=\frac{D_\sigma}{D_{\sigma}^{0}}
\label{CHI}
\end{equation}

\noindent where $D_\sigma$ is defined as the Pauli kinetic energy density and $D_{\sigma}^{0}$ is defined as the reference Pauli term for a homogeneous electron gas evaluated at the same local density. This ratio is then mapped to a range from 0 to 1 by
 
\begin{equation}
\mathrm{\eta}=\frac{1}{1+\chi^2}
\label{ELF_EQ}
\end{equation}

\noindent where $\eta$ is a dimensionless scalar quantity that is a measure related to the conditional same-spin pair probability density of finding an electron in the neighborhood of another electron with the same spin. Values close to 0 indicate a system with nearly no localized character. A value of $\eta$ = 0.5 corresponds to a degree of localization equivalent to a homogeneous electron gas, and values close to this are commonly attributed to systems with metallic character. Values close to 1 indicate a near fully localized system, such as in instances where there are bonding or lone pairs.  

\begin{figure}[H]
\centering
\includegraphics[width=1.00\textwidth]{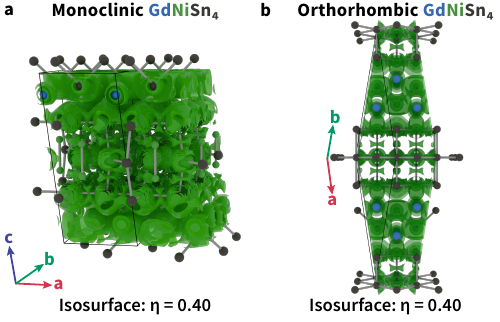}
\caption{\textbf{Electron localization function (ELF) isosurfaces for monoclinic (a) and orthorhombic (b) \ce{GdNiSn4}.} }
\label{ELF_SI_fig}
\end{figure}

Fig.~\ref{ELF_SI} shows the monoclinic and orthorhombic structures of \ce{GdNiSn4} at an equivalent isosurface value of $\eta=0.40$.

\begin{figure}[H]
    \centering
    \includegraphics[width=1\linewidth]{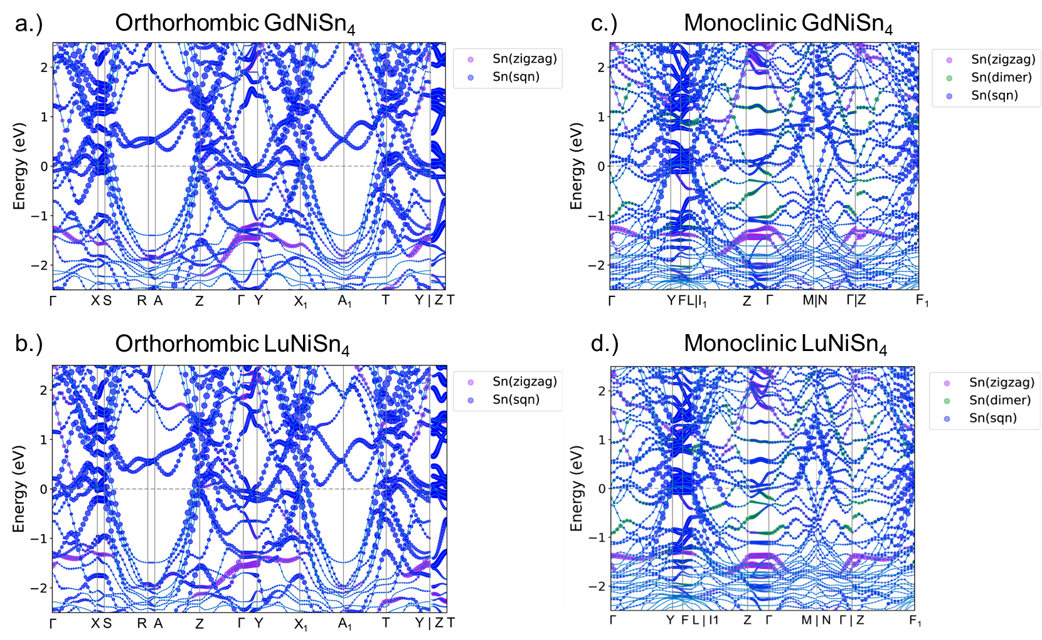}
    \caption{The projected band structures for the orthorhombic (left) and monoclinic (right) structures of \ce{GdNiSn4} (top) and \ce{LuNiSn4} (bottom). }
    \label{fig:SIPBands_EarlyVersion}
\end{figure}

\begin{figure}[H]
    \centering
    \includegraphics[width=1\linewidth]{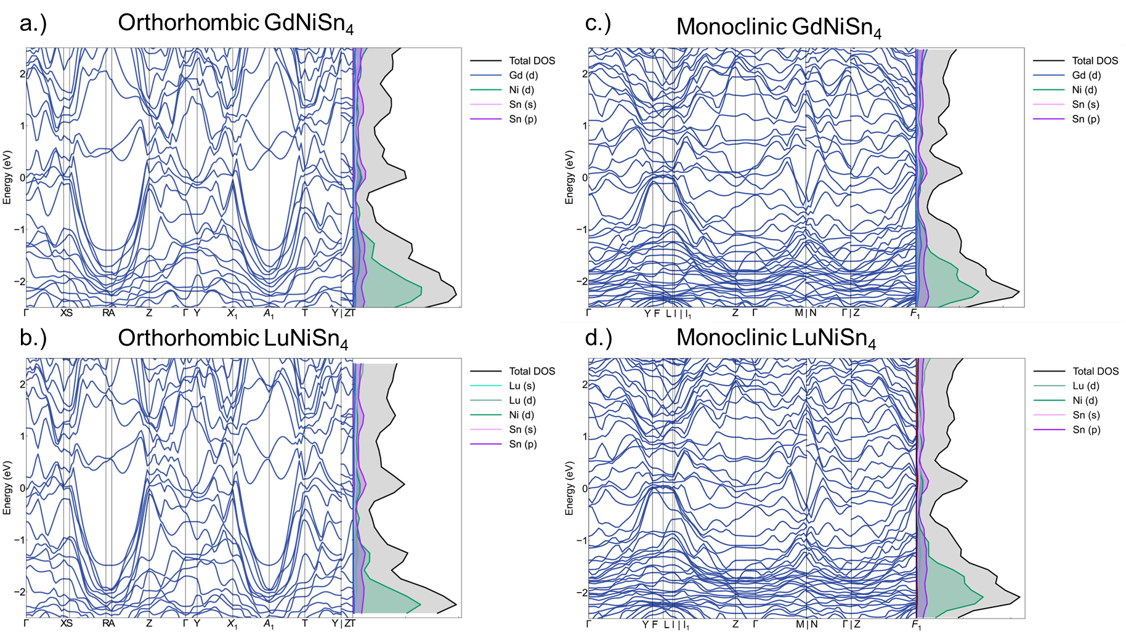}
    \caption{The bandstructures and projected density of states (PDOS) for the orthorhombic (left) and monoclinic (right) structures of \ce{GdNiSn4} (top) and \ce{LuNiSn4} (bottom).}
    \label{fig:SIBandsnDos_EarlyVersion}
\end{figure}

In Fig.~\ref{fig:SIPBands_EarlyVersion} and Fig.~\ref{fig:SIBandsnDos_EarlyVersion}, the projected band structures and density of states for both \ce{GdNiSn4} and \ce{LuNiSn4} demonstrate metallic character in both monoclinic and orthorhombic structure types. Furthermore, the states near the Fermi Level/Energy are comprised primarily of Sn atoms from the square-net in both structure types. 

\section{DFT+U Calculations}\label{DFTUSI}
\addcontentsline{toc}{section}{DFT+U Calculations}

All DFT+U calculations used here are based on the approach by Dudarev et al.~\cite{dudarev_dftu_1998}, using a Hamiltonian and total energy equation of the form shown in equations~\ref{DFT_U_Ham} and~\ref{DFT_U_Energy}. This implementation is understood as adding a positive-definite penalty functional meant to the energy expression that drives the on-site occupancy matrices towards idempotency~\cite{dudarev_dftu_1998, petersen_gdu_2006}. In the Dudarev approach, only the $U' = U-J$ value is utilized, and henceforth all variation in $U-J$ is referred to as $U'$. 

\begin{equation}
\hat{H} =
\frac{U}{2}
\sum_{m,m',\sigma}
\hat{n}_{m,\sigma}\hat{n}_{m',-\sigma}
+
\frac{(U-J)}{2}
\sum_{m \ne m',\sigma}
\hat{n}_{m,\sigma}\hat{n}_{m',\sigma}
\label{DFT_U_Ham}
\end{equation}

\begin{equation}
E^{\mathrm{DFT+U}} = E^{\mathrm{DFT}} + \frac{(U-J)}{2} \sum_{\sigma} \left[\left(\sum_{m_1} n^{\sigma}_{m_1,m_1}\right) - \left(\sum_{m_1,m_2}\hat{n}^{\sigma}_{m_1,m_2}\hat{n}^{\sigma}_{m_2,m_1}\right)\right]
\label{DFT_U_Energy}
\end{equation}

\noindent In Table~\ref{tab:DFT_U_comp}, the remaining DFT+U calculations for \ce{GdNiSn4} in either ferromagnetic or antiferromagnetic collinear magnetic structures are listed with different values of $U'$ from the Dudarev formalism for either the orthorhombic or monoclinic structure. As shown, regardless of U-level, the monoclinic form is lower in energy than the orthorhombic structure.
\begin{table}[H]
\centering
\caption{Energy comparison (eV/atom) between orthorhombic and monoclinic structures for \ce{GdNiSn4} in both ferromagnetic (FM) and antiferromagnetic (AFM) collinear magnetic structures. Bold energy values indicate the lower-energy structure for a given composition and calculation setting. A $\mathbf{k}$-point mesh of $9\times9\times9$ was used for ferromagnetic calculations and a mesh of $5\times5\times\times5$ was used for antiferromagnetic calculations. The results from varied $U'$ parameter is listed below.}
\label{tab:DFT_U_comp}
\begin{tabular}{cccccccc}
\toprule
&  \multicolumn{2}{c}{Orthorhombic (eV/atom)} & \multicolumn{2}{c}{Monoclinic (eV/atom)} \\
\cmidrule(lr){2-3} \cmidrule(lr){4-5}
 $U'$ & FM & AFM & FM & AFM \\
\midrule
0 & -5.741 & -5.777 & \textbf{-5.814} & \textbf{-5.823}\\
1 & -5.719 & -5.755 & \textbf{-5.792} & \textbf{-5.791}\\
2 & -5.700 & -5.737 & \textbf{-5.774} & \textbf{-5.773}\\
3 & -5.684 & -5.721 & \textbf{-5.758} & \textbf{-5.757}\\
4 & -5.670 & -5.707 & \textbf{-5.744} & \textbf{-5.743}\\
5 & -5.657 & -5.695 & \textbf{-5.732} & \textbf{-5.730}\\
6 & -5.645 & -5.683 & \textbf{-5.720} & \textbf{-5.719}\\
7 & -5.634 & -5.672 & \textbf{-5.709} & \textbf{-5.708}\\
8 & -5.623 & -5.662 & \textbf{-5.699} & \textbf{-5.698}\\
\bottomrule
\end{tabular}
\end{table}

\section{AI and Crystal Structure Prediction Details}\label{AI_SI}
\subsection{Generative models and CSP setting}\label{AI_SI_1}
We benchmarked two diffusion-based crystal generative models, \textit{MatterGen}~\cite{Zeni2025ADesign} and \textit{DiffCSP++}~\cite{jiao2024space}, on the task of recovering the experimental structures of \ce{GdNiSn4} and \ce{LuNiSn4}. Diffusion models define a forward noising process that progressively perturbs a crystal structure---including its lattice, fractional coordinates, and/or atomic species---toward a simple prior distribution, together with a learned reverse process that generates structures by iterative denoising. In the composition-conditioned CSP setting used here, the target composition and stoichiometry are fixed during sampling, while the model predicts lattice parameters and fractional atomic coordinates compatible with that composition. Because the experimental \ce{GdNiSn4} and \ce{LuNiSn4} structures contain 48 atoms in the conventional cell, they exceed the 20-atom cutoff of MP-20/Alex-MP-20~\cite{Jain2013,schmidt2023machine,schmidt2022dataset,cavignac2025ai}. We therefore used models trained on the MPTS-52 dataset~\cite{Jain2013}, which retains structures with up to 52 atoms. This dataset contains 40,476 entries, with 27,380 training structures, 5,000 validation structures, and 8,096 test structures. Although MPTS-52 includes larger cells than MP-20, it remains biased toward crystallographically common structures and relatively compact unit cells. As a result, models trained on MPTS-52 may still struggle on sparsely represented structure types.

\subsection{Dataset overlap analysis}\label{AI_SI_2}
Before evaluating generated structures, we checked whether MPTS-52 already contains structures equivalent to the experimental \ce{GdNiSn4} and \ce{LuNiSn4} targets. Each MPTS-52 entry was first filtered by reduced chemical formula. Only entries with the same target composition were then compared geometrically using the \textsc{pymatgen} \texttt{StructureMatcher}~\cite{ong2013python}. The matcher was used with a species-aware comparator, so only structures with identical species and composition could be accepted as matches.

We used three nested tolerance levels, denoted default, loose, and very loose, as summarized in Table~\ref{tab:levels}. These tolerances control the allowed deviations in lattice lengths, site positions, and lattice angles during structure matching. The default setting corresponds to a strict structural equivalence test, while the loose and very loose settings are intended to identify structures that preserve part of the target motif but retain non-negligible geometric distortions. 

\begin{table}[H]
  \centering
    \caption{\textbf{The three nested tolerance levels ordered tightest to loosest by the site tolerance \texttt{stol}.} \texttt{ltol} is a fractional tolerance on lattice-vector lengths, \texttt{stol} a normalized site tolerance, and \texttt{angle\_tol} a tolerance on lattice angles.}
    \vspace{6pt}
  \begin{tabular}{lccc}
    \toprule
    Level & \texttt{ltol} & \texttt{stol} & \texttt{angle\_tol} (deg)\\
    \midrule
    \texttt{default}    & 0.2 & 0.3 & 5  \\
    \texttt{loose}      & 0.3 & 0.5 & 10 \\
    \texttt{very\_loose} & 0.4 & 0.6 & 15 \\
    \bottomrule
  \end{tabular}
  \label{tab:levels}
\end{table}

For \ce{GdNiSn4}, no MPTS-52 entry has the target composition. Therefore, \ce{GdNiSn4} is absent from MPTS-52 and can be regarded as a true held-out target. For \ce{LuNiSn4}, exactly one composition-matching entry was found: \texttt{mp-31433}, which belongs to the validation split. This entry matches the experimental structure already at the default \texttt{StructureMatcher} tolerance, but it is reported in the $Cmmm$ space-group setting, No.~65, with a smaller 12-atom cell. Its residual deviations from the experimental target are:
\[
d_{\mathrm{rms}}=0.115,\qquad
d_{\max}=0.274,\qquad
\delta_{\mathrm{len}}=0.283,\qquad
\delta_{\mathrm{ang}}=6.28^{\circ},\qquad
\delta_{\mathrm{vpa}}=4.95\%.
\]
Thus \ce{LuNiSn4} is not completely absent from MPTS-52, but the matching entry is best viewed as structurally related to the target rather than an exact recovery of the same experimental cell description.

\subsection{Residual structural metrics} \label{AI_SI_3}
For every accepted match, we quantified the residual deviation from the target along two independent axes: atomic coordinates and the unit cell. Given the optimal site correspondence returned by \texttt{StructureMatcher}, we computed the normalized root-mean-square displacement

\begin{equation}
d_{\mathrm{rms}}=\sqrt{\frac{1}{N}\sum_{i=1}^{N}\left(\frac{\left\lVert\mathbf{r}^{(k)}_i-\mathbf{r}^{(\mathrm{t})}_i\right\rVert}{\ell}\right)^2}
\end{equation}
\vspace{6pt}
\noindent and the maximum paired-site displacement 

\begin{equation}
d_{\max}=\max_i\frac{\left\lVert\mathbf{r}^{(k)}_i-\mathbf{r}^{(\mathrm{t})}_i\right\rVert
}{\ell}.
\end{equation}
\vspace{6pt}

\noindent Here $N$ is the number of paired sites, $\mathbf{r}^{(k)}_i$ and $\mathbf{r}^{(\mathrm{t})}_i$ are the Cartesian coordinates of the candidate and target sites after matching, and $\ell=(V/N)^{1/3}$ is the average inter-site spacing used to normalize the displacement. Because \texttt{StructureMatcher} rescales cells to a common volume and can compare supercells, two structures may satisfy the matching criterion while still having different primitive lattice shapes or volumes. To expose these residual lattice differences, we reduced both structures to their Niggli-reduced primitive lattices and compared their sorted lattice lengths, sorted lattice angles, and volume per atom. Let ${a^{(\mathrm{t})}_j}$ and ${a^{(k)}_j}$ be the sorted primitive lattice lengths, and let ${\theta^{(\mathrm{t})}_j}$ and ${\theta^{(k)}_j}$ be the sorted primitive lattice angles. We define the following:
\begin{equation}
\begin{aligned}
  \delta_{\mathrm{len}}
    &= \max_{j}\;
       \frac{\bigl| a^{(k)}_j - a^{(\mathrm{t})}_j \bigr|}{a^{(\mathrm{t})}_j},\quad
  \delta_{\mathrm{ang}}
    = \max_{j}\;
       \bigl| \theta^{(k)}_j - \theta^{(\mathrm{t})}_j \bigr|,\quad
  \delta_{\mathrm{vpa}} = \frac{\bigl| v^{(k)} - v^{(\mathrm{t})} \bigr|}{v^{(\mathrm{t})}}, \quad
    v \equiv \frac{V_{\mathrm{prim}}}{N_{\mathrm{prim}}}.
    \label{eq:dvpa}
\end{aligned}
\end{equation}
These lattice metrics are independent of the \texttt{StructureMatcher} acceptance tolerance. In particular, $\delta_{\mathrm{vpa}}$ is computed from the unscaled primitive cells and therefore measures the genuine difference in atomic volume that is ignored by scale-invariant structure matching.

\subsection{\textit{MatterGen} benchmark} \label{AI_SI_4}
For \textit{MatterGen}, we followed the official composition-conditioned CSP setting. The composition and stoichiometric ratio were fixed, and the model generated lattice parameters and fractional coordinates. For each target composition, we generated 10,000 candidate structures using the default predictor--corrector sampler. The generated structures were compared directly to the experimental targets without further geometric relaxation. No generated structure matched either \ce{GdNiSn4} or \ce{LuNiSn4} under any of the three tolerance levels. That is, among 10,000 generated candidates for each target composition, there were no default, loose, or very loose matches.

\subsection{\textit{DiffCSP++} benchmark with crystallographic priors} \label{AI_SI_5}
We next evaluated \textit{DiffCSP++} using the pretrained MPTS-52 model. In this case, we supplied substantially stronger prior information than in the \textit{MatterGen} benchmark. Specifically, we fixed the composition and atom counts, constrained the space group to $C2/m$ (No.~12), and restricted the generation to the specific non-maximal Wyckoff positions realized in the experimental structure.

The Wyckoff constraint is important in practice. Even after fixing the composition, atom count, and space group, experimental \ce{GdNiSn4} and \ce{LuNiSn4} admit 465,335 symmetry-allowed Wyckoff templates in SG~12. Naive random sampling over these templates is therefore extremely inefficient. By imposing the experimentally realized Wyckoff template, the search is restricted to the immediate crystallographic neighborhood of the target structure.

For each target, we generated 10,000 structures and compared them to the experimental structure without further geometric relaxation. No generated structure matched either target at the default tolerance. At looser tolerance levels, \textit{DiffCSP++} generated several near matches:

\[
\begin{array}{c|ccc}
\text{Target} & \text{default} & \text{loose} & \text{very loose} \\
\hline
\ce{GdNiSn4} & 0 & 6 & 10 \\
\ce{LuNiSn4} & 0 & 13 & 18
\end{array}
\]
\vspace{6pt}
The closest generated structures had the following residual deviations:
\[
\begin{aligned}
\ce{GdNiSn4}:&\quad
d_{\mathrm{rms}}=0.108,\quad
d_{\max}=0.175,\quad
\delta_{\mathrm{len}}=0.037,\quad
\delta_{\mathrm{ang}}=15.92^{\circ},\quad
\delta_{\mathrm{vpa}}=2.87\%, \\
\ce{LuNiSn4}:&\quad
d_{\mathrm{rms}}=0.102,\quad
d_{\max}=0.162,\quad
\delta_{\mathrm{len}}=0.019,\quad
\delta_{\mathrm{ang}}=14.82^{\circ},\quad
\delta_{\mathrm{vpa}}=1.50\%.
\end{aligned}
\]
Thus the generated structures capture part of the target structural motif but remain noticeably distorted, especially in their lattice angles.

\section{Curie-Weiss Analysis}
\addcontentsline{toc}{section}{S6: Curie-Weiss Analysis}
Temperature-dependent magnetometry was performed on a \ce{GdNiSn4} single crystal up to 400K. A linear fit was performed from 350K to 400K to fit the Curie-Weiss law\cite{Mugiraneza2022Tutorial:Law}:

\begin{equation}
\chi(T) = \frac{C}{T-\theta_{\mathrm{CW}}},
\label{CW}
\end{equation}

\noindent Here, C is the Curie constant, $\theta_{CW}$ is the Curie-Weiss temperature, $\chi$ is the magnetic susceptibility, and $T$ is the temperature. The effective moment is calculated by

\begin{equation}
\mu_{\mathrm{eff}} = {\sqrt{8C}\mu_{\mathrm{B}}}
\label{exp_mu_eff}
\end{equation}

\noindent where $\mu_{\mathrm{B}}$ is a Bohr Magneton. From Hund's rules, the theoretical effective moment can be calculated as

\begin{equation}
\mu_{\mathrm{calc}} = {g_{\mathrm{j}}\sqrt{J(J+1)} \mu_{\mathrm{B}}}
\label{mu_calc}
\end{equation}

\noindent where $g_{\mathrm{j}}$ is the Landé g-factor and J is the total angular momentum\cite{Blundell2003Matter}. For a $Gd^{3+}$ free ion, $g_j$ is 2 and $J$ is $\frac{7}{2}$. This yields an effective moment of 7.94 $\mu_{\mathrm{B}}$.

\begin{figure}[H]
\centering
\includegraphics[width=1.00\textwidth]{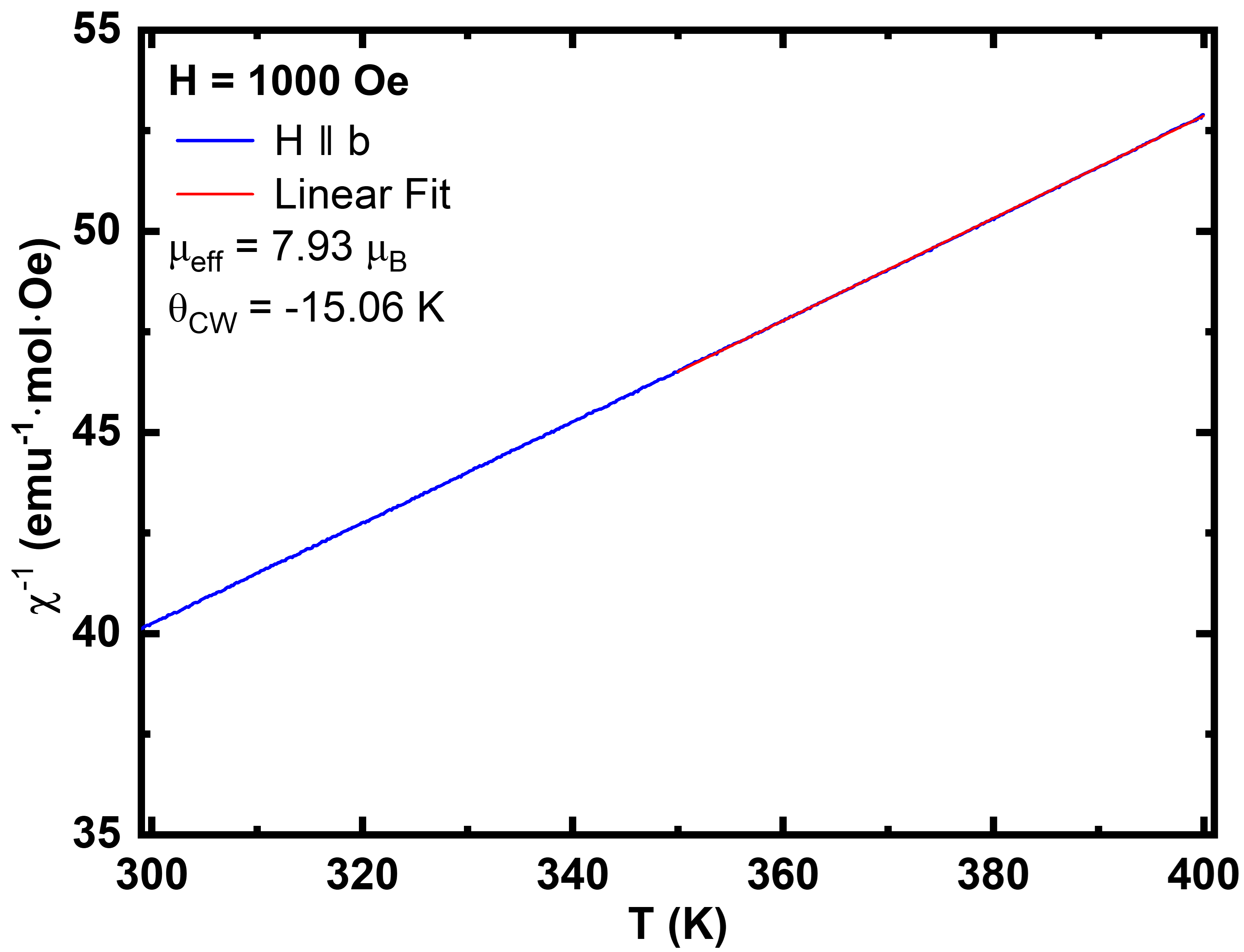}
\caption{\textbf{\ce{GdNiSn4} Curie Weiss fit.} A linear fit was performed from 350K to 400K. The experimentally derived effective moment closely matches the theoretically calculated value of 7.94 $\mu_{\mathrm{B}}$.}
\label{GdniSn4_CW}
\end{figure}

In Fig.~\ref{GdniSn4_CW}, the experimentally derived effective moment of 7.93 $\mu_{\mathrm{B}}$ closely matches the theoretically calculated effective moment of 7.94 $\mu_{\mathrm{B}}$. This suggests that Gd is the only participating magnetic ion in \ce{GdNiSn4}. 

\section{\ce{GdNiSn4} Electrical Transport Device}
\begin{figure}[H]
\centering
\includegraphics[width=1.0\textwidth]{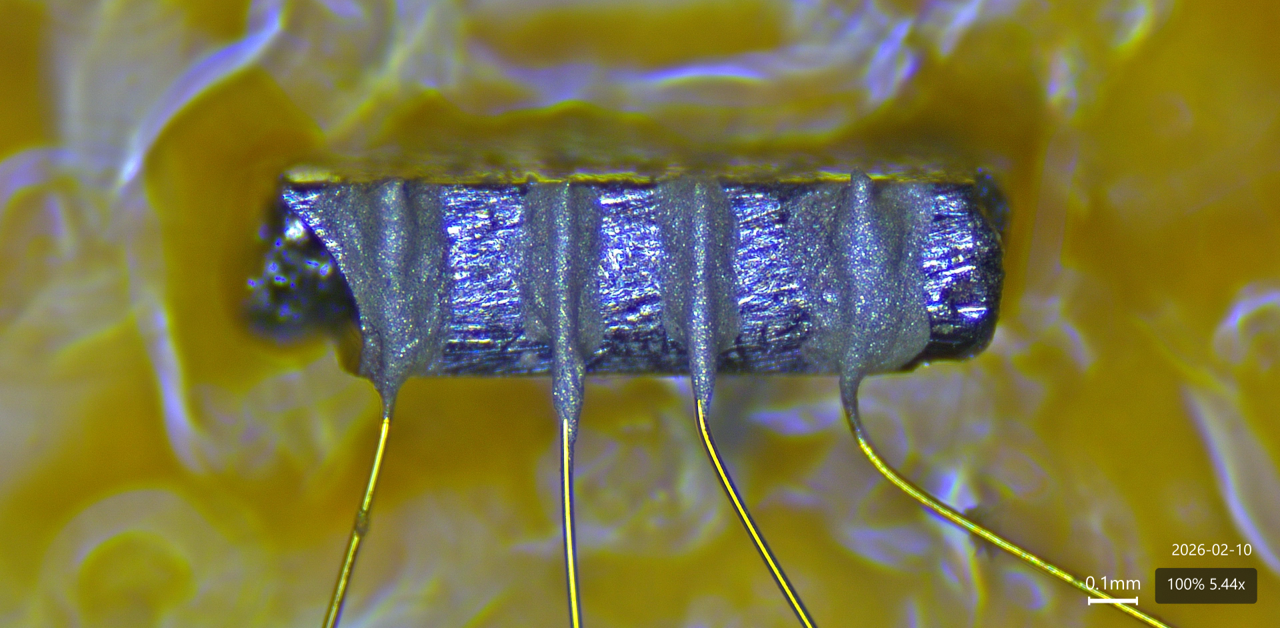}
\caption{\textbf{Electrical transport device to measure $\rho_{yy}$ in \ce{GdNiSn4}.} Gold wires were attached with conducting silver paint in a collinear four-probe configuration.}
\label{GdniSn4_ETO_Device}
\end{figure}

\addcontentsline{toc}{section}{S7: \ce{GdNiSn4} Electrical Transport Device}
Figure S\ref{GdniSn4_ETO_Device} shows a \ce{GdNiSn4} single crystal that was mechanically polished. Gold wires were configured in a collinear four-probe configuration and attached with conducting silver paint. 

\section{S8: Band structure Analysis}
\addcontentsline{toc}{section}{S8: Band structure Analysis}


\end{document}